\documentclass[showpacs,preprintnumbers,amsmath,amssymb]{revtex4}
\usepackage{graphicx}
\usepackage{dcolumn}
\usepackage{amsmath}

\newcommand{\bmf}[1]{\mbox{\boldmath$#1$}}
\newcommand{\be}{\begin{equation}}
\newcommand{\ee}{\end{equation}}
\newcommand{\ba}{\begin{eqnarray}}
\newcommand{\ea}{\end{eqnarray}}

\newcommand{\apjl}{Astrophys. J. Letters }

\newcommand{\aap}{J. Cosmo. Astropart. Phys.}
\newcommand{\simgt}{\lower.5ex\hbox{$\; \buildrel > \over \sim \;$}}
\newcommand{\simlt}{\lower.5ex\hbox{$\; \buildrel < \over \sim \;$}}

\begin{document}

\title{%
The impact of massive neutrinos on the abundance of massive clusters
}%
\author{%
 Kiyotomo Ichiki$^{1}$\footnote{E-mail address:
ichiki@a.phys.nagoya-u.ac.jp}, and 
 Masahiro Takada$^{2}$\footnote{E-mail address:
masahiro.takada@ipmu.jp} 
}%
\affiliation{%
 $^1$Department of Physics, Nagoya University, Chikusa-ku, Nagoya, 464-8602, Japan
}%
\affiliation{%
 $^2$Institute for the Physics and Mathematics of the Universe (IPMU), 
 The University of Tokyo, Chiba 277-8582, Japan
}%

\date{\today}

\begin{abstract}
We study the spherical, top-hat collapse model for a mixed dark matter
model including cold dark matter (CDM) and massive neutrinos of mass
scales ranging from $m_\nu\simeq 0.05 $ to a few $ 0.1~$eV, the range of
lower- and upper-bounds implied from the neutrino oscillation
experiments and the cosmological constraints.  To develop this model, we
properly take into account relative differences between the density
perturbation amplitudes of different components (radiation, baryon, CDM
and neutrinos) around the top-hat CDM overdensity region assuming the
adiabatic initial conditions. Furthermore, we solve the linearized
Boltzmann hierarchy equations to obtain time evolution of the
lineariezed neutrino perturbations, yet including the effect of
nonlinear gravitational potential due to the nonlinear CDM and baryon
overdensities in the late stage. We find that the presence of massive
neutrinos slows down the collapse of CDM (plus baryon) overdensity,
however, that the neutrinos cannot fully catch up with the the nonlinear
CDM perturbation due to its large free-streaming velocity for the ranges
of neutrino masses and halo masses we consider. We find that, just like
CDM models, the collapse time of CDM overdensity is well monitored by
the linear-theory extrapolated overdensity of CDM plus baryon
perturbation, smoothed with a given halo mass scale, if taking into
account the suppression effect of the massive neutrinos on the linear
growth rate.  Using these findings, we argue that the presence of
massive neutrinos of mass scales $0.05$ or 0.1~eV may cause a
significant decrease in the abundance of massive halos compared to the
model without the massive neutrinos; e.g., by 25\% or factor 2,
respectively, for halos with $10^{15}M_\odot$ and at $z=1$.
\end{abstract}
\pacs{98.80.Es,14.60.Pq,98.65.Dx}
\maketitle

\section{Introduction}
\label{sec:intro}

Galaxy clusters are the most massive, gravitationally bound objects in
the universe, therefore, their abundance and its redshift-evolution are
very sensitive to cosmology including the nature of dark energy
\cite{Vikhlininetal:09,Rozoetal:10} as well as the primordial
non-Gaussianity \cite{Dalaletal:08,Oguri:09}. There are various ongoing
and planned surveys capable of finding many clusters under a homogeneous
and well-controlled/calibrated selection; e.g., optical surveys such as
the Sloan Digital Sky Survey (SDSS) \cite{Koesteretal:07}, Subaru Hyper
SuprimeCam Survey
\cite{Miyazakietal:06}\footnote{http://sumire.ipmu.jp/en/} and the
arcminute-resolution cosmic microwave background (CMB) experiments,
Atacama Cosmology Telescope \cite{ACT:11} and South Pole Telescope
\cite{SPT:11}, with which clusters can be found  via the
Sunyaev-Zel'dovich (SZ) effect. These cluster catalogs can be used to
derive stringent constraints on cosmological parameters or more
generally test the paradigm of cold dark matter dominated structure
formation scenario \cite{Wangetal:05,Rozoetal:10,OguriTakada:11}.

The neutrino oscillation experiments have revealed that the standard
three-flavor neutrinos have non-zero masses, implying that the Big-Bang
relic neutrinos contribute to the present-day mean matter density by at
least about 0.4 and 0.8\% if the neutrinos follow the normal mass 
and inverted mass  hierarchies corresponding to the lower bounds on
the sum of neutrino masses, $0.05$ and $0.1~$eV, respectively
\citep[e.g.,][]{Takadaetal:06}. While the absolute neutrino mass scale
is still unknown, large-scale structure probes provide a powerful means
of constraining the neutrino mass
\cite{Huetal:98,AbazajianDodelson:03,Wangetal:05}. In fact the
cosmological probes have put currently the most stringent upper bound on
the sum of neutrino masses; $m_{\nu,{\rm tot}}< 0.2$ -- 0.8~eV (95\%
C.L.), the different bounds for different probes that employ 
the different level of assumptions on  nonlinear structure
formation  \cite{Ichikietal:09,Vikhlininetal:09,Saitoetal:11}. The
ongoing and upcoming cosmological surveys promise to further tighten the
neutrino mass constraint and have the potential to detect the absolute
mass scale, rather than the upper bound, if systematic errors are well
under control. See \cite{Abazajianetal:11} for the current status of the
cosmological neutrino constraints and the future prospect.

However, the previous cluster cosmology experiments rest on the use of
halo mass function calibrated based on N-body simulations that ignore
the effect of massive neutrinos (e.g.,
\cite{Tinkeretal:08,Bhattacharyaetal:11}). Although there are several
attempts to simulate nonlinear structure formation in a mixed dark
matter (CDM plus massive neutrinos) model (see
\cite{Klypinetal:93,Klypinetal:97} for the pioneer work and 
\cite{Brandbygeetal:08,BrandbygeHannestad:09,Brandbygeetal:10,Vieletal:10}
for the
recently revisited attempts),
it is still very difficult to accurately simulate the structure
formation especially for the neutrino mass scales of a few 0.1~eV or
lighter, because such light mass-scale neutrinos have a fast
free-streaming motion, larger than the gravity-induced bulk motion, at
relevant redshift and it is difficult to represent the (perturbed)
Fermi-Dirac distribution with a finite number of N-body particles at
every spatial position \cite[see][for an attempt on the grid-based
simulation of neutrinos]{BrandbygeHannestad:09}.  There are also several
attempts \cite{Saitoetal:08,Wong:08,Saitoetal:09,Lesgourguesetal:09}
aimed at developing the perturbation theory based approach to
analytically model the nonlinear structure formation for a MDM model by
extending the linear perturbation theory
\cite{MaBertschinger:95}. However, the perturbation theory based model
is only valid up to the quasi nonlinear regime and breaks down for the
nonlinear regime relevant for halo formation.

Therefore, the purpose of this paper is to develop a top-hat spherical
collapse model for a MDM model fully taking into account the effects of
multi-component system (radiation, baryon, CDM and neutrinos)
\citep[][for the pioneer work on the spherical collapse model for a CDM
model]{1969PThPh..42....9T,1972ApJ...176....1G} \citep[also
see][]{Peebles:80,Padmanabhan:93}. There are several key ingredients to
include in developing this model. Firstly, we carefully account for
differences in the density perturbation amplitudes of each component
around the top-hat CDM overdensity region assuming the adiabatic initial
conditions as predicted from standard inflation scenario. Secondly, to
achieve the desired accuracy, we solve time evolution of the density
perturbations from the deeply radiation-dominated regime to the present
time or from the sufficiently linear regime to the non-linear regime,
where the initial density perturbations are on super-horizon scales
\cite{NaozBarkana:07}. Hence, we properly take into account the
transition of perturbations from the super- to sub-horizon scales.
Thirdly, to study
the neutrino perturbations around the top-hat CDM overdensity, which
cannot be treated as a fluid, we properly solve the linearized Boltzmann
hierarchy equations \cite{MaBertschinger:95} taking into account the
nonlinear gravitational potential well due to the nonlinear CDM and
baryon perturbations.

The spherical collapse model is an approximated method for studying the
nonlinear dynamics due to the unrealistic symmetry
assumed. Nevertheless, this gives a very useful tool for studying
various effects on nonlinear structure formation; the spatial curvature
or the cosmological constant \cite{Lahavetal:91,Lilje:92}, time-varying
and/or clustered dark energy
\cite{WangSteinhardt:98,2006A&A...454...27B,Abramoetal:07,2009MNRAS.393L..31F,2010arXiv1009.0010E,Creminellietal:10},
the modified dark matter scenario \cite{2003ApJ...597..645O}, the
modified gravity scenario
\cite{SchaferKoyama:08,2010PhRvD..81f3005S,Borisovetal:11}, and the
effect of baryon perturbation \cite{NaozBarkana:07}. As for the effect
of massive neutrinos, we know that the effect is small given the current
upper bounds on the sum of neutrino masses, $\simlt$ a few 0.1~eV (at
most a few \% contribution to the matter density). Hence we expect that,
once the nonlinear dynamics is realized for a MDM model, we can
perturbatively include the effect of massive neutrino on the CDM
simulation based predictions by slightly modifying the model
ingredients. For example, the halo mass function is given as a function
of the peak height $\nu\equiv \delta_c/\sigma(M,z)$, where $\sigma(M,z)$
is the rms linear mass fluctuations of halo mass scale $M$ at redshift
$z$ and $\delta_c$ is the linear-theory extrapolated critical density as
indeed motivated by the top-hat spherical collapse model
\cite{PressSchechter:74}. Given these facts, we may be able to infer the
effect of massive neutrinos on the halo mass function once the effects
on $\delta_c$ and $\sigma(M,z)$ are realized. Thus, along this approach,
we will also discuss the impact of massive neutrinos on the abundance of
massive halos.

Throughout this paper, we employ, as our fiducial cosmological model, a
flat $\Lambda$-dominated CDM model that is consistent with the WMAP
7-year result \cite{WMAP7}; the present-day density parameters of matter
and baryon are $\Omega_{\rm m0}h^2=0.1334$ and $\Omega_{\rm
b0}h^2=0.0226$, respectively; the dimension-less Hubble parameter is
$h=0.71$; the spectral tile and normalization parameter of the
primordial power spectrum are $n_s=0.963$, and $A_s=2.43\times 10^{-9}$,
respectively.  In most parts of our paper, when adding massive neutrinos
to the fiducial cosmological model, we vary the CDM density parameter
$\Omega_{\rm c0}$ by fixing the total dark matter density to $
\Omega_{\rm c0}h^2+\Omega_{\nu 0}h^2=0.1108 $, where the energy density
parameter of massive neutrino is specified by the sum of neutrino masses
as $\Omega_{\nu0}h^2= m_{\nu, {\rm tot}}/94.1~{\rm eV}$
\cite{Takadaetal:06}. We assume the three neutrino species, and assume
one species among them is massive for simplicity.

\section{Methodology} 

In this section, we develop a method for solving a spherical top-hat
collapse of CDM overdensity
in a multi-component system, which consists of
radiation (R), baryon (b), cold dark matter (c) and massive neutrinos
($\nu$). 

\subsection{Evolutionary equation of top-hat overdensity on
superhorizon scale}

To achieve a sufficient accuracy of the spherical top-hat collapse, we
set up the initial conditions of perturbations in a deeply radiation
dominated era, $z_i=10^7$ in our case. 
The primary reason of this high initial redshift is to reproduce
the results in the earlier work \cite{NaozBarkana:07} at the limit of
massless neutrino case ($m_{\nu,{\rm tot}}\rightarrow 0$), where
\cite{NaozBarkana:07} studied the effect of baryon perturbation (without
massive neutrinos)
on halos relevant for first stars at redshift $z\simgt 30$. Note that,
for cluster-scale halo formation, we can start from the later initial
redshift such as $z\simeq 10^4$-$10^5$, but here we use the initial
redshift $z_i=10^7$ in order to retain a broader coverage of our model
validity. 
In such a radiation dominated era, the
perturbations of interest are on superhorizon scales.  To avoid gauge
ambiguities that may arise in dealing with such superhorizon-scale
perturbations, we employ the following approach.

Since we are interested in a spherical top-hat overdensity region, time
evolution of such a top-hat region can be most readily described by
a {\em perturbed} Friedman universe (e.g., \cite{Padmanabhan:93}). Here
we meant by ``perturbed'' that the top-hat region is described by a
Friedman universe with a small {\em positive} curvature that corresponds
to the top-hat density perturbation, where
the top-hat
region is embedded in the background Friedman universe that
has a flat geometry.
Thus the equation of motion for the boundary radius of top-hat
overdensity region is given as 
\begin{equation}
\left(\frac{\dot{R}}{R}\right)^2=\frac{8\pi G}{3}\left[
\bar\rho_{cb}(1+\delta_{cb})+\bar\rho_R(1+\delta_R)
\right]-\frac{k}{R^2}~,
\label{eq:pertFRW}
\end{equation}
where $R$ is the radius of the top-hat overdensity region of matter (CDM
and baryon) and radiation, $\bar\rho_{cb}$ and $\bar\rho_R$ are the mean
energy densities of matter (CDM plus baryon) and radiation, and
$\delta_{cb}$ and $\delta_R$ are their overdensities (e.g., defined as
$\delta_R\equiv \rho_R/\bar{\rho}_R-1$), respectively.  The dot notation
$\dot{~}$ denotes the time derivative, and the constant $k$ is the
effective curvature parameter $(k>0)$ which is given in terms of the
initial overdensity (see below). In this regime, massive neutrinos with
mass scales we are interested in are still relativistic and contribute
the radiation. 
Note $\delta_{cb}=(\bar{\rho}_c/\bar{\rho}_M)\delta_c+
(\bar{\rho}_b/\bar{\rho}_M)\delta_b=\delta_c $ because of
$\delta_c=\delta_b$ on superhorizon scales for adiabatic initial
conditions, where $\bar{\rho}_M$ is the mean density of total matter 
($\bar\rho_{M}=\bar{\rho}_c+\bar{\rho}_b$ or 
$\bar\rho_{M}=\bar{\rho}_c+\bar{\rho}_b+\bar{\rho}_\nu$ when   
the massive neutrino is relativistic or non-relativistic, respectively).
Also note that dark energy contribution is negligible in a
radiation dominated regime.  The scale factor for the background
universe obeys
\begin{equation}
H^2\equiv \left(
\frac{\dot{a}}{a}
\right)^2
=
\frac{8\pi G}{3}\left(\bar\rho_{cb}+\bar\rho_R\right)~.
\end{equation}

We use the linear theory predictions to determine the initial
conditions of perturbations. The spherical top-hat collapse model is
equivalent to the case that the perturbations are solved under the
synchronous gauge condition. In this case, the superhorizon-scale
perturbations grow as $\delta_{cb} \propto a^2$
\cite[][]{NaozBarkana:07,MaBertschinger:95}. Assuming this growing mode
and using the mass conservation $R^3\bar{\rho}_{cb}(1+\delta_{cb}
)={\rm constant}$ yield the initial condition for the velocity
$\dot{R}/R$: 
\begin{equation}
\left.\frac{\dot{R}}{R}\right|_i=H_i-\frac{2}{3}\frac{\delta_{cb,
 i}}{1+\delta_{cb, i}}H_i,
\label{eq:iniR}
\end{equation}
where $H_i\equiv H(t_i)$ and $\delta_{cb,i}\equiv \delta_{cb}(t_i)$. By inserting
Eq.~(\ref{eq:iniR}) into Eq.~(\ref{eq:pertFRW}) at the initial time
$t_i$ we can express the effective curvature parameter $k$ in terms of
the initial overdensity $\delta_{cb, i}$ as
\begin{eqnarray}
\frac{k}{R_i^2}&\equiv &\frac{8\pi
 G}{3}\left(
\bar\rho_{cb,i}+\frac{4}{3}\bar\rho_{R,i}
\right)\delta_{cb,i} +H_i^2
 \left[
\frac{4}{3}\frac{\delta_{cb, i}}{1+\delta_{cb,i}}
-\frac{4}{9}\left(\frac{\delta_{cb,i}}{1+\delta_{cb,i}}\right)^2
\right],
\label{eq:superhorizon}
\end{eqnarray}
where we have used the adiabatic initial condition to re-express
radiation perturbation in terms of matter perturbation as
$\delta_R\simeq (3/4)\delta_{cb}$.
Hence Eq.~(\ref{eq:pertFRW}) can be re-written as
\begin{eqnarray}
\left(\frac{\dot{R}}{R}\right)^2&=&\frac{8\pi G}{3}\left[
\bar\rho_{cb}(1+\delta_{cb})+\bar\rho_R\left(
1+\frac{4}{3}\delta_{cb}
\right)
\right]-\frac{k}{R^2}.
\label{eq:tophat_sh}
\end{eqnarray}
The relation between
$R$ and $\delta$ follows from the mass conservation; 
$(1+\delta_{cb})=
(R/R_i)^{-3}
(1+\delta_{cb,i})(a/a_i)^3$.
Hence, given the initial conditions on $R_i$ (or $\delta_{cb, i}$) and
$\dot{R}_i$, Eq.~(\ref{eq:tophat_sh}) can be solved numerically to
obtain time evolution of the top-hat overdensity $\delta_{cb}$ 
until
 the top-hat region
enters into the horizon.

Eq.~(\ref{eq:tophat_sh}) is an exact equation that can be applied even
if the perturbation amplitude is large (unrealistic though). 
In the radiation dominated regime, the linear theory gives a good
approximation.
By linearizing Eq.~(\ref{eq:tophat_sh}) we can derive the differential
equation which governs time evolution of the linear perturbation
$\delta^{\rm L}$:
\begin{equation}
H \dot{\delta}^{\rm L}_{cb}=-4\pi G\left(\bar{\rho}_{cb}
+\frac{4}{3}\bar{\rho}_R
\right)\delta^{\rm L}_{cb}+\frac{16\pi
G}{3}\frac{a_i^2}{a^2}\bar{\rho}_{cb,i}\delta_{cb,i}, 
\label{eq:linear_sh}
\end{equation}
where we have used the fact $\bar{\rho}_{R,i}\gg \bar{\rho}_{M,i}$
at the initial
redshift. Again note $\delta_{cb}\approx \delta_{cb}^{L}$ to a good
approximation in this regime. 

\subsection{Evolutionary equations of  perturbations on sub-horizon scales}

When the overdensity region enters into the horizon, perturbations
of different components evolve in different ways; the CDM overdensity
continues to grow, and baryon and neutrinos cannot grow together with
CDM. We below describe our treatments of each
component's dynamics on subhorizon scales after the horizon crossing.

\subsubsection{CDM perturbation}

CDM plays a major role in the spherical collapse model. When the top-hat
overdensity region enters into the horizon, we use the following
equation, obtained in the Newtonian gauge, in order to solve the
dynamics up to the nonlinear collapse of spherical CDM overdensity region:
\begin{equation}
\frac{\ddot{R}_{c}(t)}{R_{c}(t)}=-\frac{4\pi G}{3}
\left[
\bar{\rho}_{\rm tot}(t)+\bar{P}_{\rm tot}(t)
\right]
-
\frac{G \delta\! M(<R_{c};t)}{R_{c}^3(t)},
\label{eq:tophat_cdm}
\end{equation}
where $R_c(t)$ is the radius of the top-hat region of CDM overdensity,
and $\bar{\rho}_{\rm tot}$ and $\bar{P}_{\rm tot}$ denote the mean
energy and pressure densities, which determine the cosmic expansion
history over the range of radiation, matter and dark energy dominated
eras.  Note $R_c(t_{\rm enter})=R(t_{\rm enter})$ at the horizon
crossing, where $R(t_{\rm enter})$ is the radius of the initial top-hat
overdensity region discussed in the preceding section.
The quantity $\delta M$ is the mass fluctuation within the
CDM-overdensity sphere, and includes contributions from CDM, baryon and
neutrino perturbations: $\delta\!  M\equiv \delta\!  M_c+\delta\!
M_b+\delta\!  M_\nu$. Note that we ignore perturbations of dark energy
in this paper \cite[see][for the spherical collapse model with dark
energy perturbations, 
but without massive neutrinos]{Abramoetal:07,2010arXiv1009.0010E}.

For CDM perturbation, 
the mass conservation within the top-hat
region holds: 
\begin{eqnarray}
M_c &=&\frac{4\pi
R_c^3}{3} \bar{\rho}_c(t) [1+\delta_c(t)] 
=\frac{4\pi
R_{c,i,{\rm enter}}^3}{3}\bar{\rho}_{c,i,{\rm enter}}(1+\delta_{c,i,{\rm
enter}} )
\simeq \frac{4\pi R_{c,i}^3}{3}\bar{\rho}_{c,i}, 
\label{eq:enclosed_mass}
\end{eqnarray}
where $\delta_c(t)$ is the overdensity at time $t$, and the quantities
with subscript ``${}_{c,i,{\rm enter}}$'' denote their quantities at
the horizon crossing.
For the perturbations
of interest, the horizon crossing is earlier than the decoupling epoch, where
the perturbations are well in the linear regime.
In the equation above
we have used $\delta_{c,i}\ll 1$ at the initial redshift, so 
used the fact $\bar{\rho}_{c,i}(1+\delta_{c,i})\simeq \bar{\rho}_{c,i}$.
We determine
$\delta_{c,i,{\rm enter}}$, $R_{c,i,{\rm enter}}$ and $\dot{R}_{c,i,{\rm
enter}}$ by matching the values to those from
Eq.~(\ref{eq:tophat_sh}) at the horizon crossing. We again stress that,
by matching  these initial conditions
from superhorizon to subhorizon scales, we can achieve a
sufficiently accurate setup of the initial conditions needed for the nonlinear
spherical collapse dynamics, even for a high collapse-redshift such as
$z\simgt 30$.

To solve Eq.~(\ref{eq:tophat_cdm}), we need to specify the mass
fluctuations of baryon and massive neutrinos, $\delta M_b(t)$ and
$\delta M_\nu(t)$, within the spherical region of radius $R_c(t)$ at
each time step. However, unlike CDM, the mass fluctuations are not
conserved within the CDM top-hat region, because baryon is dragged out of the CDM potential
well due to the 
tight
coupling with photons before the decoupling epoch,
and neutrinos are free-streaming out of the CDM potential well due
to their large thermal velocities \footnote{See
http://cmb.as.arizona.edu/\~eisenste/acousticpeak/acoustic\_physics.html
for a pedagogical illustration of density evolutions of different
components around the CDM density perturbation peak in the linear
regime.}. In fact the linear perturbation theory gives $\delta_c\gg
\delta_b \gg \delta_\nu$ at the decoupling epoch. Hence, for simplicity,
we assume $\delta_b=\delta_\nu=0$ during epochs after the horizon
crossing to the decoupling epoch. One can then find that 
Eq.~(\ref{eq:tophat_cdm}) gives growing modes of
$\delta_c\propto \ln a$ and $\propto a$ in the radiation and matter
dominated eras in the linear regime (i.e. $\delta_c\ll 1$), as expected
from  the linear perturbation
theory.

After the decoupling epoch $(z\simeq 10^3)$, baryon becomes
cold, and follows the mass conservation. We will below describe our
treatments of baryon and neutrino perturbations in subsequent
subsections.

Linearizing Eq.~(\ref{eq:tophat_cdm}) yields the following equation to
describe the evolution of linear CDM perturbation:
\begin{equation}
\ddot{\delta}^{\rm L}_{c}+2{ H}\dot{\delta}^{\rm
 L}_{c}-4\pi G\left[
\bar\rho_c\delta^{\rm L}_c+\bar\rho_b\delta^{\rm
			   L}_b(<R^{\rm L}_c)
+\bar\rho_\nu \delta^{\rm L}_\nu(<R^{\rm L}_{c})\right]=0,
\end{equation}
where $\delta^{\rm L}_{c}$ is the linear CDM overdensity,
and $\delta^{\rm L}_b(<R_c)$ and $\delta^{\rm L}_\nu(<R_c)$ are the
linear density perturbations averaged within the sphere of radius
$R_c^{\rm L}$. Here the comoving radius of $R_c^{\rm L}$ is set to the
same in the linear theory: $
R_c^{\rm L}/a=R^{\rm L}_{c,i}/a_{i}$. 
The initial conditions for $\delta^{\rm L}_{c,i}$ and $\dot{\delta}^{\rm
L}_{c,i}$ are set by matching to Eq.~(\ref{eq:linear_sh}) at the horizon
crossing. Before the decoupling epoch and until sufficiently higher
redshift before halo formation, the equation above gives a good
approximation to Eq.~(\ref{eq:tophat_cdm}). 

\subsubsection{Baryon perturbation}

Next let us consider evolution of baryon perturbation in a CDM top-hat
overdensity region. 

After the horizon-crossing of the spherical top-hat overdensity region
until the decoupling epoch ($z_{\rm dec}$), baryon is tightly coupled to
photon and the baryon density perturbation cannot grow. More exactly,
the baryon-photon fluid oscillates according to the acoustic sound wave
in the CDM potential well  -- the baryon acoustic
oscillations (BAO) \cite{HuSugiyama:95}. The characteristic scale of this
clustering is about 150~Mpc for our fiducial cosmological model.
Therefore, even if the baryon perturbation initially had a top-hat
overdensity profile as in CDM, the baryon perturbation becomes increasingly
spatially-extended than the CDM top-hat region as time goes by until
$z_{\rm dec}$. The linear perturbation theory predicts that the baryon
density perturbation becomes much smaller in the amplitude than the CDM
density perturbation at the decoupling epoch
for length scales of interest; $\delta_b\ll
\delta_c$ at $z_{\rm dec}$ for scales of interest.

Therefore we simply assume that the baryon density perturbation averaged
within the CDM top-hat region is $\delta_b(<R_c)=0$ during epochs from
the horizon-crossing to the decoupling epoch; $z_{\rm enter}>z\ge z_{\rm
dec} $, where $z_{\rm dec}$ is specified once a background cosmological
model is given, e.g., from the CAMB code \cite{CAMB}.
 We have checked that the spherical collapse of CDM
overdensity is not changed even if we instead use the different
assumption; $\delta_b(<R_c)=\delta_b(<R_c; z_{\rm enter})$, where
$\delta_b(<R_c; z_{\rm enter})$ is the baryon overdensity when the CDM
top-hat region had the horizon-crossing.

After the decoupling epoch baryon becomes a cold component, and can
cluster together with CDM perturbation. We can thus use the spherical
collapse equation to solve nonlinear evolution of baryon perturbation in
the CDM top-hat region just like the CDM case
(Eq.~[\ref{eq:tophat_cdm}]):
\begin{equation}
\frac{\ddot{R}_{b}(t)}{R_{b}(t)}=-\frac{4\pi G}{3}
\left[
\bar{\rho}_{\rm tot}(t)+\bar{P}_{\rm tot}(t)
\right]
-
\frac{G \delta\! M(<R_{b};t)}{R_{b}^3(t)}.
\label{eq:tophat_baryon}
\end{equation}
Here $R_b(t)$ is the radius of baryon overdensity region, which is
chosen from the radius satisfying $R_b=R_c$ at $z_{\rm dec}$.
After the decoupling epoch, the mass conservation holds:
$R_b(t)^3\bar{\rho}_b(t) \left[1+\delta_b(<R_b)\right]={\rm constant}$.
The initial conditions are set to $\dot{R}_b=HR_b$ at $z_{\rm dec}$,
which corresponds to $\dot{\delta}_b=0$, motivated by the fact that
baryon-photon coupling prevents baryon perturbation from growing before
the decoupling epoch. Thus the baryon velocity perturbation is different from
that of CDM perturbation, given as $\dot{R}_{b}> \dot{R}_{c}$, so the
time evolution of baryon radius $R_b(t)$ differs from the CDM radius
$R_c(t)$. As times goes by, the baryon perturbation eventually catches
up with the CDM perturbation as we will explicitly show below.

\subsubsection{Neutrino perturbation}
\label{sec:neutrino}

The neutrino perturbation on sub-horizon scales cannot be captured by
the CDM potential well due to the large free-streaming velocity. As a
result, the neutrino perturbation around the CDM overdensity extends out
to radius comparable with the free-streaming scale that is given as $
\lambda_{\rm fs}\simeq a^{-1}H^{-1}\sigma_{v,\nu}(z)$ in the units of
comoving scale; here $\sigma_{v,\nu}$ is the velocity dispersion of the
Fermi-Dirac distributed neutrinos (see Appendix A in
\cite{Takadaetal:06} for the definition). For example, 
for neutrinos of mass scale $m_\nu\simeq
0.1$~eV,
$\lambda \simeq
20~h^{-1}$Mpc at $z=0$. 
To model these physical processes we use the modified CAMB code
to solve time evolution of neutrino clustering, where we properly take
into account the nonlinear gravitational potential due to nonlinear CDM
and baryon perturbations in the late stage (see \cite{Saitoetal:08} for
the similar approach to solving the evolution of mildly nonlinear
perturbations).

To be more precise, we solve the linearized Boltzmann equation that
governs time evolution of the neutrino distribution function
$f(x^i,q,\hat{n}_j,\tau)=f_0(\epsilon)[1+\Psi(x^i,q,\hat{n}_j,\tau)]$. 
Here $\tau$ is the conformal time, 
$q$ and
$\hat{n}_j$ denote the comoving momentum and its direction, $\epsilon=(q^2+a^2
m_\nu^2)$ is the proper energy times scale factor $a(t)$, 
$f_0$ is
the background distribution function (the Fermi-Dirac distribution) and $\Psi$ is
the perturbed distribution function.
The Boltzmann equation in an expanding universe can be reduced to the
following hierarchical equations in Fourier space
\cite{MaBertschinger:95}:
\begin{eqnarray}
 \dot\Psi_0 &=& -\frac{qk}{\epsilon}+\frac{1}{6}\dot\phi\frac{d\ln f_0}{d\ln
  q}~,\nonumber \\
  \dot\Psi_1 &=& \frac{qk}{3\epsilon}(\Psi_0-2\Psi_2)-\frac{\epsilon
   k}{3q}\psi \frac{d\ln f_0}{d\ln  q}~,\label{eq:boltz-hierarchy}\\
 \dot\Psi_\ell&=&\frac{qk}{(2\ell+1)\epsilon}\left[
\ell\Psi_{\ell-1}-(\ell+1)\Psi_{\ell+1}
\right]\hspace{2em} (l\ge 2)~,\nonumber
\end{eqnarray}
where $\psi$ and $\phi$ are the metric perturbations in the Newtonian
gauge, and 
the perturbed distribution function $\Psi$ is expanded in terms of the
Legendre polynomials as 
\begin{equation}
\Psi(\vec{k},\hat{n},q,\tau)\equiv \sum_{\ell=0}^{\infty}(-i)^\ell(2\ell+1)\Psi_\ell(\vec{k},q,\tau)P_\ell(\hat{k}\cdot
 \hat{n}).
\end{equation}

Our main interest is to study the impact of massive neutrinos on the
spherical collapse of CDM overdensity in the matter
or dark energy dominated era.
At redshifts after the decoupling epoch, 
the difference between metric perturbations $\psi$
and $\phi$, which arises from anisotropic stress, is negligible:
$\psi=\phi$.
In this case,
the potential perturbation is given by the Poisson equation as
\begin{equation}
-k^2 \psi(k,\tau)=4\pi G a^2 \sum_{i={\rm b,c,}\nu} 
\bar{\rho}_i(\tau) \tilde{\delta}_i(k,\tau),
\end{equation}
where $\tilde{\delta}_i(k)$ denotes the Fourier-transformed
coefficients of the $i$-th component ($i=$c, b, $\nu$). 
We insert the nonlinear top-hat overdensities of CDM and baryon into the
Poisson equation above to compute the potential including the nonlinear
contribution. Note that we take the center of the top-hat region as the
coordinate center, which makes the potential dependent only on the
length of wavevector $\bmf{k}$; $\psi(\bmf{k})=\psi(k)$. 
The corresponding potential for the linear-theory
extrapolated density perturbations is computed from 
$-k^2 \psi^{\rm L}(k,\tau)=4\pi G a^2 \sum_{i={\rm b,c,}\nu}
\bar{\rho}_i(\tau) \tilde{\delta}^{\rm L}_i(k,\tau)~$.  Since we have assumed
that CDM and baryon overdensities take spherical top-hat profiles, whose
radii are $R_c$ and $R_b$, respectively, the Fourier-transformed
counterparts of CDM and baryon perturbations are given analytically as
\begin{equation}
\tilde{\delta}_{i}(k,\tau)=\frac{4\pi}{k^3}\left[
\sin(kR_{i})-kR_{ i}\cos(kR_{ i})
\right]\delta_{ i}(\tau),
\label{eq:f_cdm_tophat}
\end{equation}
where $i=$c or b, and
 $\delta_{\rm c,b}(\tau)$ are the mean overdensities of CDM and
baryon within their respective top-hat regions. 
We can compute the neutrino perturbation in Fourier
space from the zero-th moment of the 
perturbed distribution function:
\begin{equation}
\tilde{\delta}_{\nu}(k,\tau) = 4\pi a^{-4} \int q^2 dq \epsilon f_0
 \Psi_0(k,\tau). 
\label{eq:nu_dens}
\end{equation}
The corresponding linear perturbation is given as 
$\tilde{\delta}^{\rm L}_{\nu}(k,\tau) = 4\pi a^{-4} \int q^2 dq \epsilon f_0
 \Psi^{\rm L}_0(k,\tau)$,
a standard output of the CAMB code. 

We also need to compute the real-space overdensity of neutrino
perturbations at each time step, which is needed for the
spherical collapse model of CDM and baryon overdensities. The radial
profile of neutrino density perturbation and its mass fluctuation within
the CDM or baryon top-hat regions are
\begin{eqnarray}
\delta_\nu(r; \tau) &=& \int_0^\infty 4\pi k^2 dk \frac{\sin(kr)}{kr}
\tilde{\delta}_\nu(k,\tau)~,\\
\delta M_\nu(<R_{c,b}; \tau)&\equiv& \int_0^{R_{c,b}}4\pi r^2 dr \bar{\rho}_\nu
 \delta_\nu(r), 
\end{eqnarray}
We use the publicly available code FFTLog \cite{2000MNRAS.312..257H} to
compute the density profile at each time step, because the code allows
for a fast computation of the integration involving the Bessel
function kernel. Similarly, the density
profile and the mass fluctuation for the linear neutrino perturbation
can be obtained by using the linear density perturbation
$\tilde{\delta}^{\rm L}_\nu(k,\tau)$ instead of $\tilde{\delta}_\nu(k,\tau)$. 

We employ the decoupling epoch $z_{\rm dec}$ to set up the initial
conditions of neutrino perturbations as in the baryon perturbations.
Since the neutrino perturbations at $z_{\rm dec}$ are well in the
linear regime, we can determine the initial conditions by matching to
the linear perturbation theory predictions. To be more precise, assuming
the adiabatic initial conditions, we can determine the neutrino density
perturbation at $z_{\rm dec}$ around the top-hat CDM overdensity
region \footnote{Again also see {\sf
http\://cmb.as.arizona.edu/\~{}eisenste/acousticpeak/acoustic\_physics.html}
for the similar method for the linear perturbations}:
\begin{equation}
\tilde{\delta}_\nu(k,\tau_{\rm dec})=\delta_\nu^{\rm L}(k,\tau_{\rm
 dec})=\frac{T_\nu(k,\tau_{\rm dec})}{T_{c}(k,\tau_{\rm dec} )}\delta^{\rm L}_{c}(k,\tau_{\rm dec}), 
\label{eq:nu_ini}
\end{equation}
where $\tilde{\delta}_{c}(k,\tau_{\rm dec})$ is the Fourier transform of
the CDM top-hat overdensity (Eq.~[\ref{eq:f_cdm_tophat}]), and the
functions $T_\nu(k,\tau_{\rm dec})$ and $T_{c}(k,\tau_{\rm dec})$ are the
transfer functions of massive neutrinos and CDM at $z_{\rm dec}$,
respectively. 
We use the CAMB outputs to obtain the transfer functions.  The ratio
$T_\nu/T_c$ takes into account the relative amplitude difference of
neutrino and CDM perturbations at $z_{\rm dec}$ under the adiabatic
initial conditions. 
We compute the zero-th moment of the perturbed distribution $\Psi_0(k,z_{\rm
dec})$ at the initial time by 
multiplying the CAMB output ${\rm Tr}[\Psi_0^{\rm
L}(k,\tau_{\rm dec})]$ with $\tilde{\delta}_c(k,z_{\rm
dec})/T_{c}(k,\tau_{\rm dec})$ so that Eq.~(\ref{eq:nu_dens}) gives the
neutrino density perturbation around the CDM top-hat
overdensity, where ${\rm Tr}[\Psi_0^{\rm
L}(k,\tau_{\rm dec})]$ is the linear transfer of the zero-th moment of
the perturbed distribution function. 
 Similarly, we compute the higher-order function
$\Psi_l(k,z_{\rm dec})$ ($l\ge 1$) from the CAMB outputs ${\rm Tr}[\Psi^{\rm
L}_l(k,z_{\rm dec})]$ multiplied by $\tilde{\delta}_c(k,z_{\rm
dec})/T_{c}(k,z_{\rm dec})$. We can obtain the subsequent evolution of
neutrino perturbations by solving the Boltzmann equation hierarchies
Eq.(\ref{eq:boltz-hierarchy}) given these initial conditions at $z_{\rm dec}$.

Our approach above is still an approximation; we used the linearized
Boltzmann equations where each term in the hierarchies depends linearly
on perturbation quantities. In other words, even though we include the
effect of nonlinear gravitational potential, we ignore nonlinear terms in
the Boltzmann equations, e.g., the term proportional to $O(\Psi \phi)$,
which can be important if the neutrino perturbation itself becomes
nonlinear. We will come back to this issue later. 

\subsection{Summary: our recipe for solving the spherical top-hat 
 collapse model}

\begin{table*}
\begin{tabular}{l|lclcl}\hline
 & superhorizon && subhorizon before the decoupling ($z_{\rm dec}$)
 &\hspace*{2em}& subhorizon after $z_{\rm dec}$\\  \hline
 CDM& perturbed FRW (Eq.~\ref{eq:tophat_sh}) 
&$\Longleftrightarrow$ &sph. collapse model 
(Eq.~\ref{eq:tophat_cdm})
&& sph. collapse model (Eq.~\ref{eq:tophat_cdm})
\\
baryon& perturbed FRW (Eq.~\ref{eq:tophat_sh}) && $\delta_b=0$ 
&
& sph. collapse model (Eq.~\ref{eq:tophat_baryon}) 
  \\ 
massive-$\nu$& perturbed FRW (Eq.\ref{eq:tophat_sh}) & & $\delta_\nu=0$&$\Longleftrightarrow$&
 linearized Boltzmann eqs. (Sec.~\ref{sec:neutrino}) \\ 
radiation & perturbed FRW (Eq.\ref{eq:tophat_sh}) & & -- && --\\ \hline
\end{tabular}
\caption{A quick summary of our recipe used for solving the spherical
 top-hat collapse model in a multi-component system 
(CDM, baryon, massive
neutrinos and radiation), where the initial top-hat CDM overdensity 
drives the
 collapse at late times. 
 We set the initial time to be in the deeply
radiation dominated regime, $z_i= 10^7$, in order to achieve a
sufficient accuracy of setting up the initial conditions needed for
the nonlinear dynamics. Also note that we assumed
the adiabatic initial conditions which determine the relations between
density perturbation amplitudes of different components at the initial
time (see text for details). The density perturbations of interest are
 on superhorizon scales at the initial time, enter into the horizon and
 evolve on subhorizon scales. 
The notation
``$\Longleftrightarrow$'' denotes the matching of perturbations between
different equations. Massive neutrinos with mass scales ($\simlt
\mbox{a few }0.1~$eV) become non-relativistic after the
decoupling epoch $(z_{\rm dec})$, and we set up the neutrino perturbation
amplitudes at $z_{\rm dec}$ by matching with the CAMB outputs (see text
 for details).
\label{tab:recipe}  }
\end{table*}
Here is a quick summary of the procedures we take for solving the
spherical top-hat collapse model in a multi-component system of CDM,
baryon and neutrinos. 
\begin{enumerate}
\item Choose a target mass scale of halo, $M_c$, to determine the
      comoving scale of the spherical top-hat CDM overdensity region
      via the relation
      $R_{c,0}(M_c)=(3M_c/4\pi \bar{\rho}_{c,0})^{1/3}$.
\item Solve the linear CDM perturbation of the comoving scale $R_{c,0}$
      from the initial time $z_i= 10^7$ to the present time based on the
      linear perturbation theory, assuming the adiabatic initial
      conditions.
In these calculations we
      properly take into account the fact that the density perturbations
      are on super-horizon scales in early epochs, enter into the
      horizon, and evolve on sub-horizon scales.
\item Choose a target collapse redshift $z_{\rm coll}$ that
      corresponds to the halo formation. Then, as a first guess,
 normalize the initial top-hat CDM overdensity, $\delta_c(z_i)$, for a
      given cosmological model in such a way that the linear-theory
      extrapolated overdensity satisfies the condition $\delta^{\rm
      L}_c(z_{\rm coll})= 1.686(1+z_{\rm coll})$, a prediction for the
      collapse redshift for Einstein de-Sitter model, a CDM model
      without baryon and massive neutrino contributions.
\item Solve the spherical top-hat collapse of CDM overdensity
      (Eq.~[\ref{eq:tophat_cdm}]), coupled with the spherical top-hat
      collapse of baryon overdensity (Eq.~[\ref{eq:tophat_baryon}]) and
      the linearized Boltzmann equations of neutrino perturbations (see
      Sec.~\ref{sec:neutrino}).
\item Solve the nonlinear evolution of CDM overdensity until
      $\delta_c\rightarrow \infty$. Iteratively solve the spherical
      collapse model by changing the initial overdensity amplitude
      $\delta_c(z_i)$ until the CDM overdensity becomes to collapse at
      the target collapse redshift, $\delta_c(z_{\rm coll})\rightarrow
      \infty$. Also obtain the linear-theory extrapolated overdensity
      for CDM or CDM plus baryon perturbations, 
      $\delta^{\rm L}_c(z_{\rm coll})$ or $\delta^{\rm L}_{cb}(z_{\rm
      coll})$.
\end{enumerate}
Table~\ref{tab:recipe} also gives a quick summary of these treatments,
clarifying which equations we use for solving the spherical top-hat
collapse model.

It would also be useful to explicitly list the assumptions we employ for
the spherical collapse model:
\begin{itemize}
\item We used the {\em linearized} Boltzmann hierarchy equations to solve the
      time evolution of {\em linearized} neutrino
      perturbations. However, we include the effect of nonlinear
      gravitational potential due to the nonlinear CDM and baryon
      density perturbations. 
\item We assumed the top-hat profiles of CDM and baryon perturbations.
\item We set the baryon and neutrino density perturbations to zero,
      i.e. $\delta_b=\delta_\nu=0$, before the decoupling epoch, because
      we found it gives a good approximation compared to a more rigorous
      calculation under the adiabatic initial conditions.
\end{itemize}
For the second assumption above, rigorously speaking, even if we
consider a spherically symmetric top-hat overdensity region at the
initial time, the radial profile become changed by the presence of
radiation, baryon and neutrino perturbations, which go out of the CDM
overdensity region after the horizon crossing. As a result the
overdensity region no longer obeys a top-hat profile. However, a
spherical top-hat collapse model is anyway an approximated method for
studying the nonlinear dynamics of the initial overdensity regions,
preferentially representing density peaks in the primordial
perturbations. Hence the top-hat overdensity region can be interpreted
as the average density contrast around such density peaks after making a
top-hat filtering with the smoothing scale of target halo mass scale
\citep[e.g.,][for a similar discussion]{Padmanabhan:93}. For these
reasons, our treatment of assuming a spherically symmetric, top-hat
overdensity for CDM perturbation is adequate enough for our purpose. Our
main goal is to study the impact of massive neutrinos on the spherical
collapse by comparing the results with and without massive neutrino
contribution. Also note that our method includes the limit of spherical
collapse for a pure CDM model without baryon and massive neutrino
contributions, by imposing $\Omega_{\rm b}=\Omega_\nu=0$.

\section{Results}

\subsection{Spherical collapse in a mixed dark matter model}

\begin{figure}[t]
\begin{center}
\includegraphics[width=0.95\textwidth]{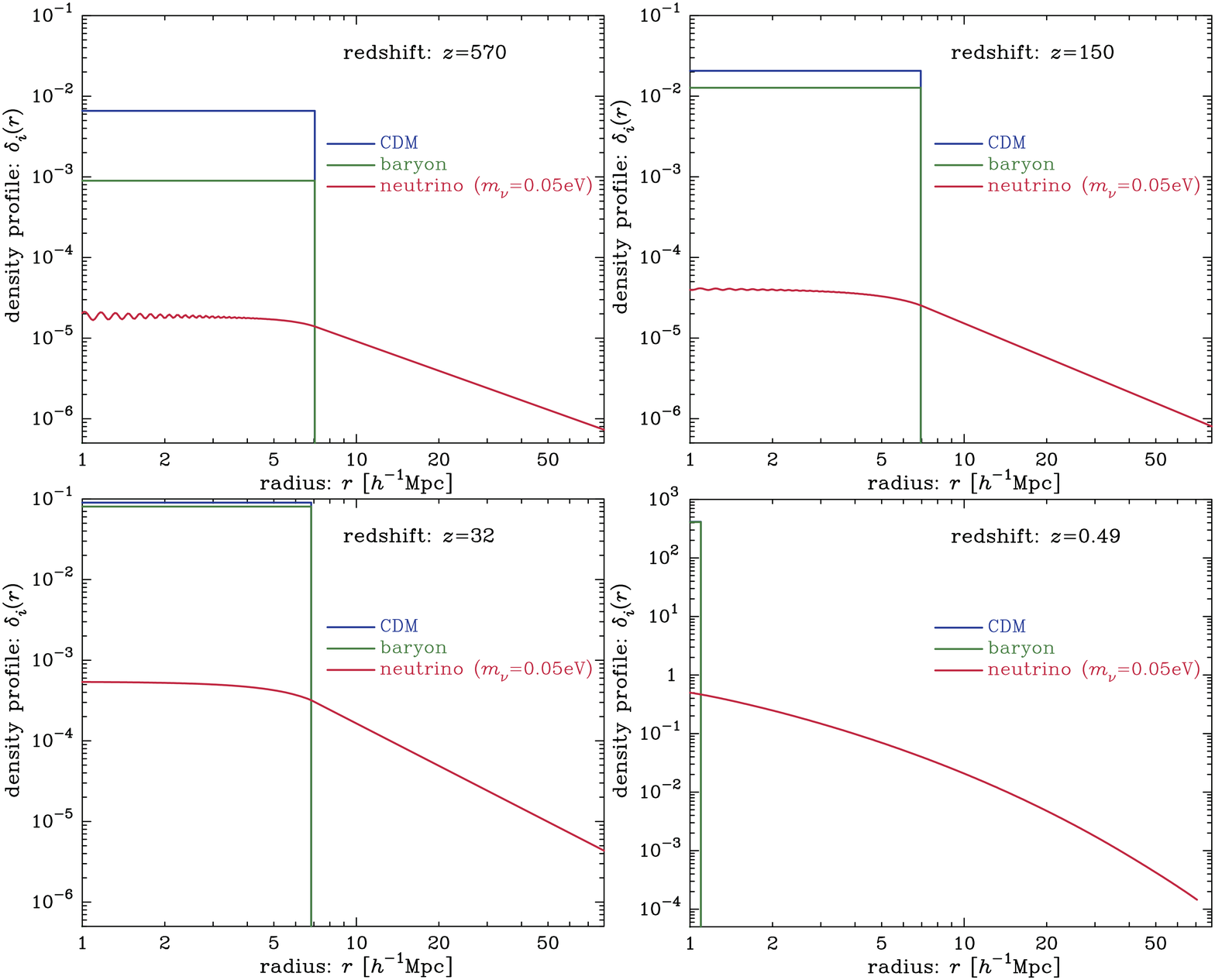}
\end{center}
\vspace*{-2em}
\caption{Radial profile of density perturbation for each component (CDM,
baryon and massive neutrino) in the CDM top-hat overdensity region
assuming the adiabatic initial conditions to determine relative
amplitudes of different components. We plot the profiles in units of
comoving scale, ${\rm Mpc}/h$.  We assumed $m_\nu=0.05~{\rm eV}$ for
neutrino mass scale, $R=6.89~h^{-1}{\rm Mpc}$ for radius of the top-hat
region, which corresponds to halo mass scale $M=10^{14}~h^{-1 }M_\odot$
for our fiducial cosmological model, and determined the initial density amplitude so
that the top-hat region collapses around $z=0.5$.  Note that, for the
baryon perturbation, we show its density contrast within the CDM top-hat
region for illustrative clarity (the baryon top-hat perturbation
computed is spatially more extended, and holds the mass conservation
within its own top-hat region).  The different panels show the profiles
at different redshifts as indicated.  At sufficiently early redshift
such as $z=570$, after the decoupling epoch, $\delta_{\rm
b},~\delta_{\rm \nu}\ll \delta_{\rm c}$, because baryon was coupled with
radiation until the decoupling epoch ($z\simeq 1100$) and neutrino was
free-streaming out of CDM potential well.  As time goes
by, the baryon perturbation eventually catches up with the CDM
perturbation. Then at redshifts lower than $z\simeq 30$ for this case,
the top-hat radius starts to shrink and the top-hat dynamics deviates
from the linear theory and enters into the nonlinear regime.  As for
neutrinos, the large velocity dispersion of neutrino particles prevents
from catching up with the nonlinear collapse and then becomes to have a
more spatially-extended profile than the CDM and baryon top-hat region.
Even when CDM and baryon collapse ($z\simeq 0.5$), the neutrino
perturbation stays in the quasi nonlinear regime as $\delta_\nu \lesssim
1$.}  \label{fig:sp_collapse}
\end{figure}

To compute the spherical collapse model, we assume a flat-geometry cold
dark matter, $\Lambda$ dominated cosmological model ($\Lambda$CDM) that
is consistent with the WMAP results \cite{WMAP7}. We further need to
specify neutrino parameters.  In this paper, we assume standard three
flavors of neutrinos. Since structure formation is sensitive only to the
sum of the three-species neutrino masses, we assume, for simplicity,
that only one species of neutrinos are massive and other two species are
massless. In this case, the neutrino free-streaming scale is shortest
for a fixed total neutrino mass (or a fixed $\Omega_{\nu 0}$), and
therefore the neutrino has the largest ability to cluster on small
scales compared to a case that the total neutrino mass is split into
different species. We then study how the spherical collapse is affected
by massive neutrino assuming the mass scale ranging from 0.05 to a few
0.1~eV. This range of mass scales covers the lower and upper bounds on
the neutrino mass implied from the neutrino oscillation experiments and
the cosmological constraints
\citep{Ichikietal:09,Saitoetal:11,Abazajianetal:11}. In most parts of
this paper, when we add the massive neutrinos, we keep the energy
density of ``dark matter'' ($\Omega_{\rm c0}+\Omega_{\nu0}$) and other
parameters fixed, and vary $\Omega_{\rm c0}$.

Fig.~\ref{fig:sp_collapse} shows how the CDM top-hat overdensity grows
as a function of cosmic time.  We assumed 
$m_\nu=0.05~{\rm eV}$ for neutrino mass, a mass scale close to the lower bound implied
from the normal mass hierarchy,
and halo mass scale $M=(4\pi /3)R_{c,i}^3\bar{\rho
}_{c,i}=10^{14}h^{-1}M_\odot$ corresponding to the comoving radius,
$R_{c,0}=6.89~h^{-1}$Mpc, for our fiducial $\Lambda$CDM model.  By
considering massive halos, we can estimate the largest effect
of neutrinos on the spherical top-hat collapse, as such a massive halo
has the ability to trap neutrinos around it due to the deepest
gravitational potential well.
We, as a working example, set the initial CDM overdensity so that the
top-hat region collapses at redshift $z_{\rm coll}\simeq 0.5$ for a model
without massive neutrino.

The plot also shows the baryon density contrast within the CDM top-hat
region as well as the radial profile of neutrino perturbations out to
radii outside the top-hat region. Note that, for the baryon
perturbation, we show only its density contrast within the CDM top-hat
region (more exactly speaking, the baryon overdensity region is slightly
more spatially-extended than the CDM as we described above). At
sufficiently high redshifts such as $z=570$, the baryon and neutrino
perturbations are smaller in the amplitudes than the CDM density
contrast. Then at lower redshifts, the baryon perturbation eventually
catches up with the CDM perturbation. For this particular case, 
at lower redshifts
$z\simlt 30$, the comoving radius of CDM top-hat region starts to
shrink, entering into the nonlinear regime (or equivalently deviating
from the linear evolution). The figure explicitly demonstrates that the
CDM and baryon perturbations, i.e. cold components, can collapse
together having $\delta_c, \delta_b\rightarrow \infty$ at the collapse
redshift.

On the other hand, the neutrinos of this mass scale cannot catch up with
the CDM perturbation due to the large free-streaming velocity. To be
more precise, the present-day free-streaming scale in the comoving scale
unit is $\lambda_{\rm fs}\simeq \sigma_{v,\nu}H_0^{-1}\sim 40~h^{-1}{\rm
Mpc}$ (see Appendix~A in \cite{Takadaetal:06}), which is much larger
than a few Mpc, a scale of the virial radius of massive halos. The plot
shows that the neutrinos are indeed clustering around the CDM top-hat
region, and become to have the radial profile. The neutrino perturbation
peaks at the center of the CDM top-hat region, but the density contrast
is still smaller than unity, so not yet in the highly nonlinear
regime. More precisely, the neutrino perturbation averaged within the
CDM top-hat region $\delta_\nu \simeq 0.19$ at the collapse redshift.
We have also checked that, when neutrino mass is in the range smaller
than a few 0.1~eV, the neutrino density contrast grows only up to the
weakly nonlinear regime $\delta_\nu \simeq $ a few even at the collapse
time (for the case of $m_{\nu,{\rm tot}}=0.1~$eV, $\delta_\nu\simeq
0.51$ at the collapse redshift). Note that, for the halo of
$10^{15}M_\odot$ and the collapse redshift $z_{\rm coll}\simeq 0.5$,
$\delta_\nu\simeq 0.54$ and $2$ for $m_{\nu,{\rm tot}}=0.05$ and
$0.1~$eV, respectively).  Therefore our approach using the linearized
Boltzmann equations for neutrino perturbations is approximately
validated.  Once the halo is formed via virialization of the kinetic and
gravitational bound energies, the neutrino would become stably clustered
around the halo region as studied in
\cite{SinghMa:03,Abazajianetal:05}. Nevertheless the resulting neutrino
overdensity is much smaller than the CDM and baryon perturbations in a
halo region, therefore we ignore the neutrino mass contribution to the
halo mass in the following analysis for simplicity.

\begin{figure}[t]
\begin{center}
\includegraphics[width=0.6\textwidth]{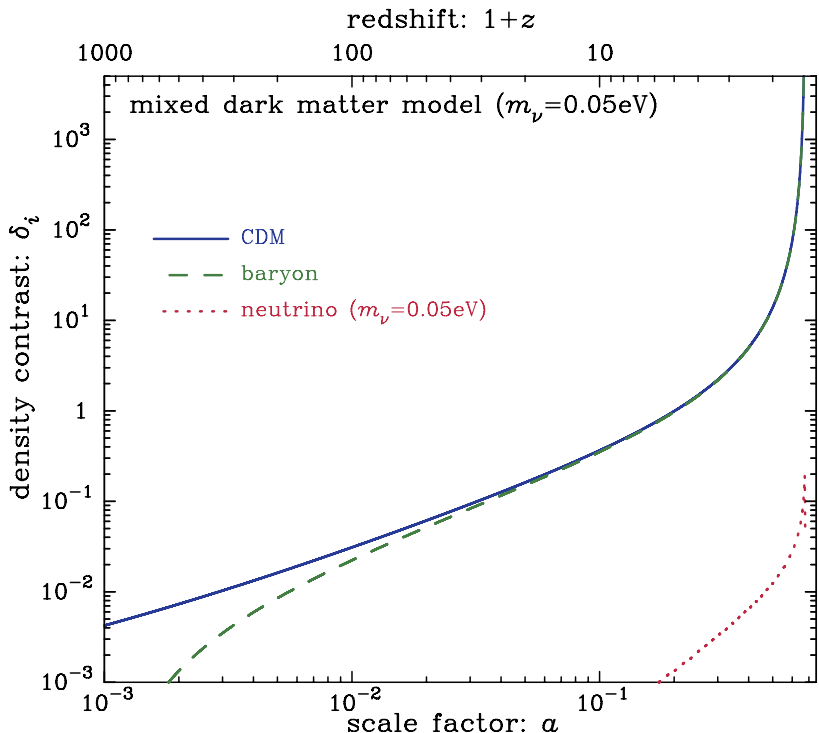}
\end{center}
\vspace*{-2em}
\caption{Time evolution of the density contrast of each component,
 averaged within the CDM top-hat region. We used the same initial
 conditions and model parameters as in Fig.~\ref{fig:sp_collapse}.
For this particular case, 
the neutrino perturbation averaged within the CDM top-hat region
$\delta_\nu\simeq
 0.19$ at the collapse redshift, and therefore is
 still in the quasi nonlinear regime. 
}
 \label{fig:density}
\end{figure}

Fig.~\ref{fig:density} shows the time evolution of density contrasts
within the top-hat region for each component. Note that we computed the neutrino
density contrast by averaging the density profile within the top-hat
region. 
The CDM and
baryon collapse at $z_{\rm coll}\simeq 0.48$ having
$\delta_c,\delta_b\rightarrow \infty$. The neutrinos are affected by the
nonlinear clustering of CDM perturbations, but do not enter into the
highly nonlinear regime.

\begin{figure}
\begin{center}
\includegraphics[width=0.6\textwidth]{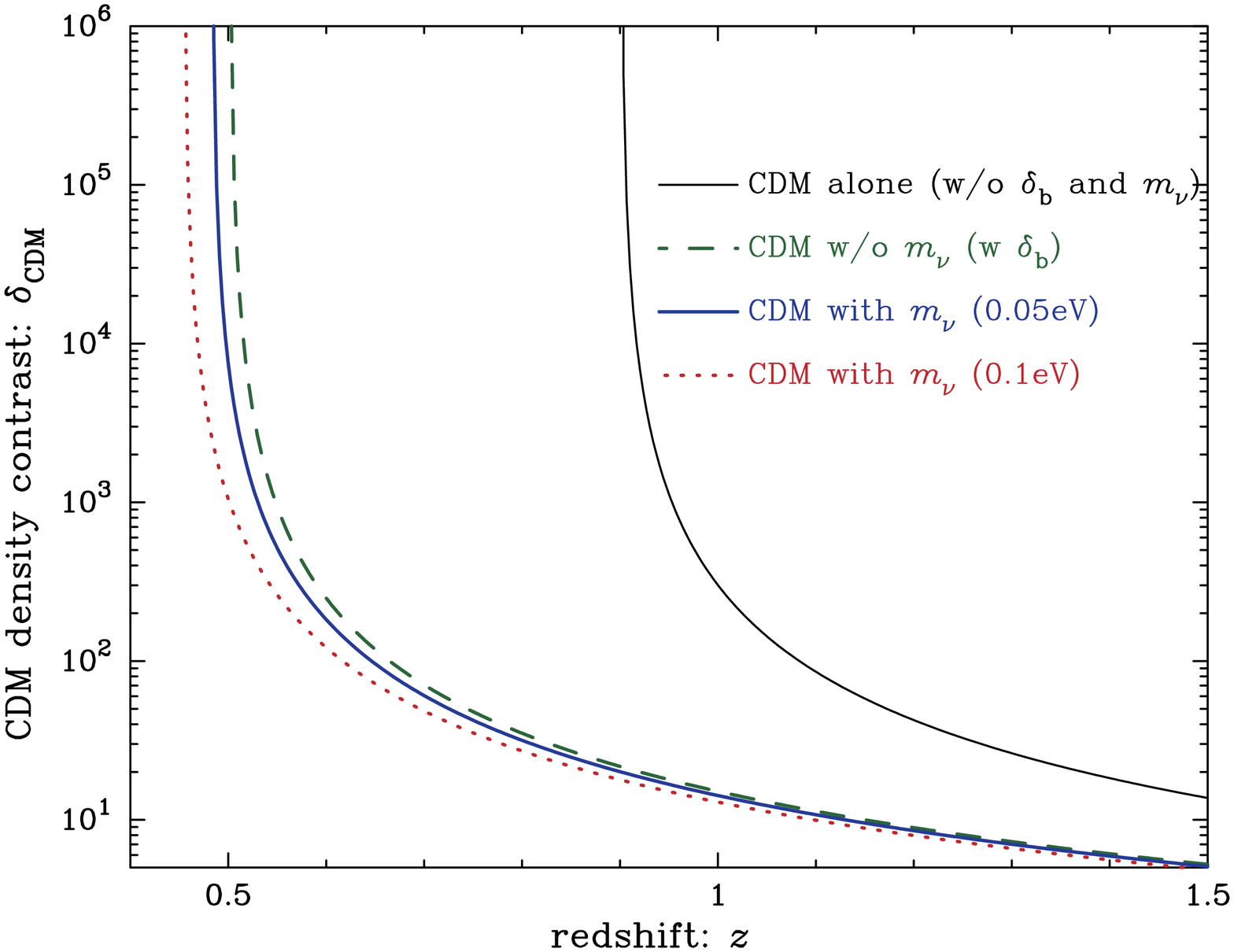}
\end{center}
\caption{Time evolution of the CDM top-hat overdensity for different
models. The dashed curve shows the result for our fiducial cosmological
model without massive neutrino. We again assumed the halo mass scale
$M=10^{14}~h^{-1}M_\odot$ and determined the initial CDM perturbation so
that the top-hat region collapses at $z\simeq 0.5$.  The solid curve
shows the result when ignoring the baryon perturbation, where the CDM
perturbation amplitude is set so as to match that of the dashed curve at
the decoupling epoch $z\simeq 1100$. Note that the model of the solid
curve leads the linear-theory extrapolated overdensity to be
$\delta^{\rm L}=1.686$ at redshifts before dark energy domination in the
cosmic expansion (the result shown here is affected by dark energy
domination). Comparing the solid and dashed curves manifests that the
presence of baryon perturbation, which has a smaller amplitude at
earlier redshifts as implied from Figs.~\ref{fig:sp_collapse} and
\ref{fig:density}, delays the spherical collapse. The solid and dotted
curves show the results when further including massive neutrino for a
fixed dark matter density $\Omega_{\rm c0}+\Omega_{\nu0}$. The
neutrino perturbation is smoother than that of CDM perturbation, and
delays the spherical collapse.  } \label{fig:compare_womnu}
\end{figure}

In Fig.~\ref{fig:compare_womnu}, we compare the spherical collapses of
CDM perturbation for models with and without baryon perturbation and/or
massive neutrino contributions. 
We used the
same initial conditions of CDM perturbation except for the result
without baryon perturbation (labelled as ``w/o $\delta_b$ and
$m_\nu$'').  For the model without baryon perturbation (thin solid
curve), we set the CDM perturbation amplitude to match the CDM amplitude
at $z_{\rm dec}$ for our fiducial model (without massive neutrino), but
set $\Omega_{\rm m0}=\Omega_{\rm c0}$. In this case, 
if we set a sufficiently early collapse
redshift, this model gives the collapse redshift given by
$\delta_c(z_{\rm coll})\simeq 1.686(1+z_{\rm coll})$, the case for an
Einstein de-Sitter model. At later collapse redshift, the cosmological
constant  becomes dominant in the cosmic
expansion, and the collapse redshift differs from the Einstein de-Sitter
prediction.  This result is compared with other curves. First, the dashed
curve shows the collapse of CDM perturbation $\delta_{\rm c}$ when the
baryon perturbation is added. The presence of baryon perturbation, which
cannot grow during epochs from the horizon enter until the decoupling
epoch, delays the spherical collapse. The delay is rather significant,
from $z_{\rm coll} \simeq 0.9$ to $0.5$, and therefore the result
suggests that we need to carefully take into account the effect of
baryon perturbation on the nonlinear structure formation in N-body
simulations, especially when we set up the initial conditions (we will
discuss this issue later).  The bold solid and dotted curves show the
spherical collapse when massive neutrino is further added for mass
scales of $0.05$ and $0.1$~eV, respectively. The collapse redshifts are
further delayed as $z_{\rm coll}\simeq 0.48$ and 0.45, respectively.
These mass scales correspond to the lower bounds on total neutrino mass
for the normal and inverted mass hierarchies.  Thus, adding the
smoother, massive components into the CDM perturbation progressively
delays the spherical collapse.

\begin{figure}[t]
\begin{center}
\includegraphics[width=17cm,height=7cm]{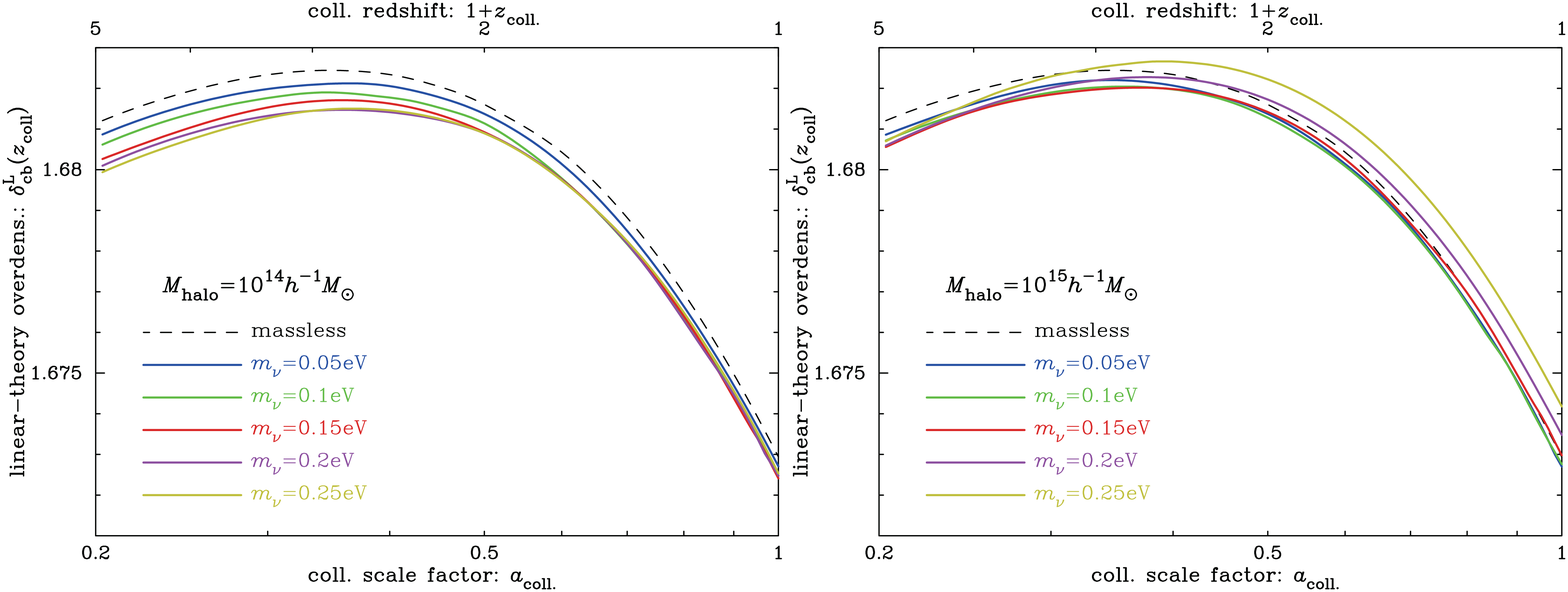}
\end{center}
\caption{The linear-theory extrapolated density contrast of CDM plus
 baryon perturbation at the collapse redshift -- the so-called critical
 density that can be used to infer the collapse redshift based on the
 linear theory. The left and right panels show the results for halo mass
 scales $M=10^{14}$ and $10^{15}~h^{-1}M_\odot$, respectively. The
 different curves are the results without and with massive neutrino
 contribution assuming different neutrino mass scales. Note that the
 density contrast shown here is not for CDM perturbation alone, and the
corresponding critical density of  CDM perturbation is greater than
shown in this plot. 
The overall change in the critical density from the Einstein de-Sitter result
$\delta^{\rm L}=1.686$
arises from the effect
 of baryon perturbation for higher redshifts, 
while the change at lower redshift $z\simlt
 1 $ is due to dark energy domination in the cosmic expansion. The
 effect of massive neutrino is in the range of the different curves. The
 curves  show non-trivial dependence on neutrino mass scale (see the
 next figure). 
}  \label{fig:dlc}
\end{figure}
\begin{figure}[t]
\begin{center}
\includegraphics[width=0.6\textwidth]{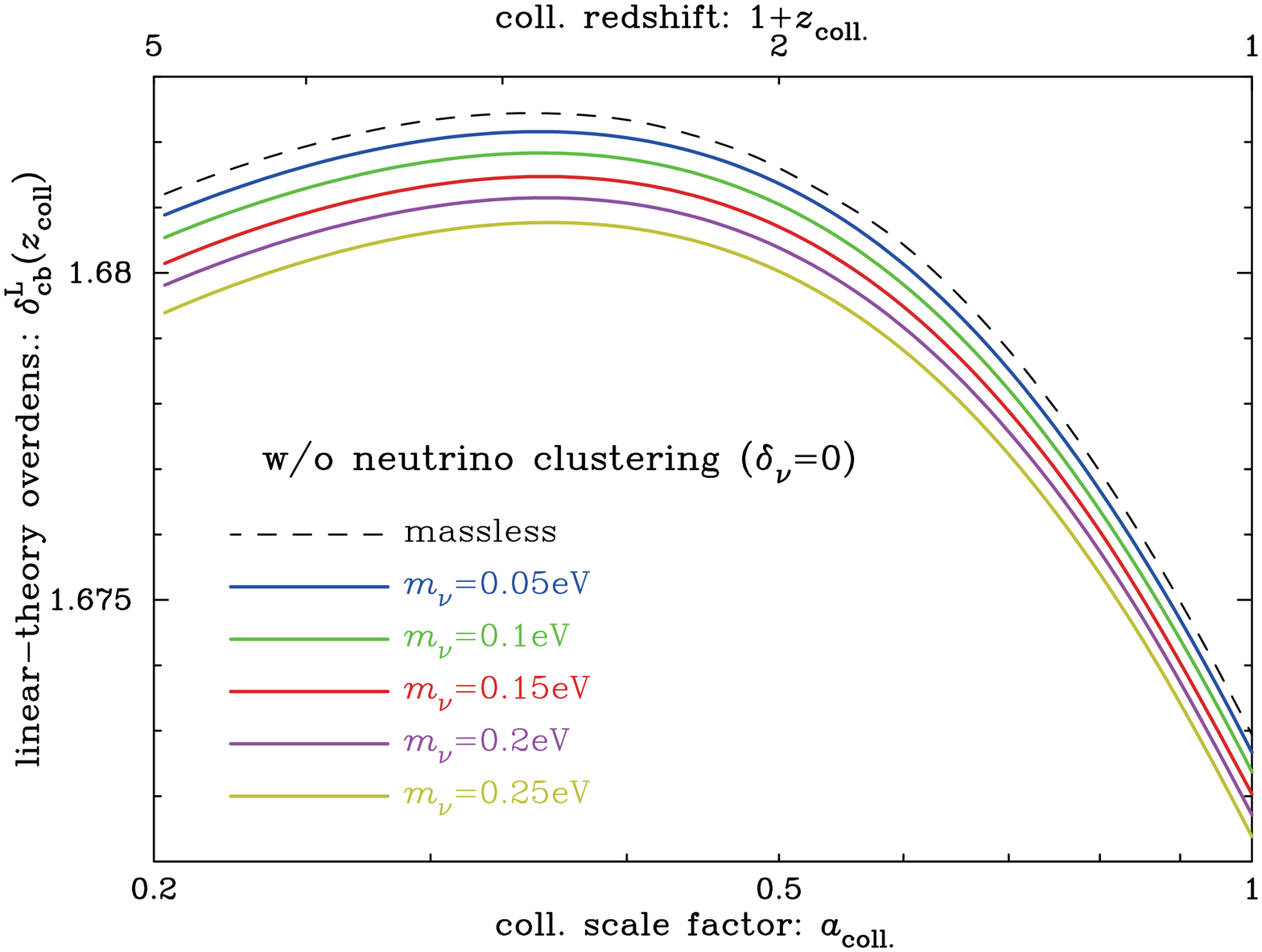}
\end{center}
\caption{Similarly to the previous figure, but we here ignored the
 effect of neutrino perturbation on the spherical collapse and on the
 linear density calculation; i.e. we
 set $\delta_\nu=0$ and $\delta^{\rm L}_\nu=0$. 
The critical overdensity becomes smaller at each
 redshift with increasing the neutrino mass scale. Therefore, comparing
 this figure with Fig.~\ref{fig:dlc} clarifies that the non-trivial
 dependence of $\delta^{\rm L}_{cb}$ on neutrino mass scale is due to
 the effect of neutrino perturbation (see text for details). }
\label{fig:dlc_wodnu}
\end{figure}

We then study the linear-theory extrapolated overdensity at the collapse
redshift (we will often call it the critical overdensity hereafter). In
Fig.~\ref{fig:dlc} we show the critical overdensity of CDM plus baryon
perturbation, $\delta^{\rm L}_{cb}(z_{\rm coll};R)$, smoothed with
length scales $R=6.89$ and $14.8~h^{-1}{\rm Mpc}$ corresponding to halo
mass scales $M=10^{14}$ and $10^{15}~h^{-1}M_\odot$, respectively (the
subscript ``$cb$'' stands for CDM plus baryon).
The overdensity $\delta^{\rm L}_{cb}(z; r)$ can serve as a
clock to infer the collapse redshift, because it can be easily computed once
the initial power spectra of CDM and baryon perturbations, halo mass
scale and
cosmological model are specified.  In an Einstein de-Sitter
universe ($\Omega_{\rm c0}=1$), which includes CDM alone, the critical
overdensity can be derived analytically
\cite{1969PThPh..42....9T,1972ApJ...176....1G} and is found to be
$\delta^{\rm L}_c(z_c) = 1.686$, independently of halo mass and collapse
redshift.  Even for a model with curvature or dark energy contribution,
the critical density $\delta^{\rm L}_c(z_c)\approx 1.686$ to a good
approximation,  independently
of halo mass \cite{Lahavetal:91,Lilje:92,WangSteinhardt:98}.
The top solid curve shows the critical overdensity when baryon is
included, but massive neutrino is ignored. The critical overdensity
differs from 1.686, and the change of $\delta_{cb}^{\rm L}(z_{\rm coll};
R)$ is due to the presence of baryon perturbation, which has a smaller
amplitude than the CDM perturbation at higher redshifts (e.g., see
Fig.~\ref{fig:sp_collapse}). Our result is consistent with the result in
\cite{NaozBarkana:07}, where they studied the effect of baryon
perturbation on the spherical collapse at high redshift for much smaller
halos that are relevant for first stars. Although the presence of
smoother baryon perturbation delays the spherical collapse, it leads to
the {\em smaller} $\delta^{\rm L}_{cb}(z_{\rm coll}; R)$ than 1.686.
However, note that the
linearly-extrapolated overdensity for CDM perturbation alone,
$\delta^{\rm L}_c(z_{\rm coll}; R)$, is indeed greater than 1.686. That
is, the initial CDM top-hat overdensity with greater amplitude than
expected from $\delta_c^{\rm L}(z_{\rm coll};R)=1.686$ is needed so that
it collapses at a given collapse redshift $z_{\rm coll}$. The curve
peaks around $z_{\rm coll}\simeq 2$ ($a_{\rm coll}\simeq 0.33$) 
 having $\delta^{\rm L}_{cb}(z_{\rm coll}; R)\simeq
1.682$.
At the lower redshifts than $z_{\rm
coll}\simeq 2$, especially at $z_{\rm coll}\simlt 1$, 
the critical overdensity becomes smaller than the peak
value
 due to the effect of the cosmological constant. When the
cosmological constant or more generally dark energy becomes to dominate
the cosmic energy density, the accelerating cosmic expansion slows down
the growth of CDM plus baryon perturbation, and delays the spherical
collapse. This yields the smaller critical overdensity. Both the linear
and nonlinear growths of CDM plus baryon perturbation are delayed by the
cosmic acceleration, and the linear growth is more suppressed than the
nonlinear growth, because the spherical collapse eventually separates
from the cosmic expansion in the nonlinear stage, and becomes more
affected by the self-gravity of nonlinear CDM plus baryon
overdensity. For these reasons, when the growth of density perturbations
is suppressed by the faster cosmic expansion than the Einstein de-Sitter
model, it generally leads to the smaller critical overdensity
$\delta^{\rm L}_{cb}(z_{\rm coll}; R)$ than 1.686.

The other curves in Fig.~\ref{fig:dlc} show the results for $\delta^{\rm
L}_{cb}(z_{\rm coll}; R)$ when including the massive neutrino for a
fixed total matter density $\Omega_{\rm m0}$.  The presence of massive
neutrino further delays the spherical collapse (see
Fig.~\ref{fig:compare_womnu}), and in turn leads to the smaller critical
overdensity $\delta^{\rm L}_{cb}(z_{\rm coll}; R)$, the same trend for
the effects of baryon perturbation and the cosmological
constant. However, the massive neutrino only decreases $\delta^{\rm
L}_{cb}(z_{\rm coll}; R)$ by less than 0.1\% compared to the solid
curve, for these halo mass and neutrino mass scales and over a range of
redshifts we have studied. This small change in $\delta^{\rm
L}_{cb}(z_{\rm coll}; R)$ can be contrasted with the effect on the
linear growth rate; the growth rate is suppressed by the amount of $\sim
4f_\nu$ at relevant redshift compared to the growth rate without the
massive neutrino \cite{Huetal:98,Takadaetal:06}, corresponding to 1.6
and 3.2\% suppression for the neutrino mass scales of 0.05 and 0.1~eV,
respectively. The results imply that the neutrino effect on the
spherical collapse is well captured by the linear growth rate of CDM
plus baryon perturbation. 

The different curves show that the change in $\delta^{\rm L}_{cb}$ is
not monotonic with changing neutrino masses, when keeping the
present-day dark matter
density $(\Omega_{\rm c0}+\Omega_{\nu0})$ fixed. This non-trivial dependence
can be understood as follows.  
The neutrino effect on the spherical collapse arises from its effect on the
cosmic expansion history and the gravitational collapse of CDM
perturbation.  First, the presence of massive neutrino leads to a faster
cosmic expansion during the neutrino was relativistic, which slows down
the growth of CDM perturbation. Note that the neutrino becomes
non-relativistic when $T_{\rm CMB}\simeq m_{\nu}$.  Secondly, the
neutrino perturbation does contribute to the nonlinear gravitational
collapse of CDM perturbation, and therefore accelerates the spherical
collapse to some extent. The net effect arises from these competing
effects.  We can study which of these two effects is more important as
follows. Fig.~\ref{fig:dlc_wodnu} shows the results when we ignore the
neutrino perturbation ($\delta_\nu=0$) in both the spherical collapse
calculation as well as the linear growth calculation for CDM plus baryon
perturbation. The figure shows that, with increasing the neutrino mass
scale, the spherical collapse more delays and the critical overdensity
becomes monotonically smaller. Hence, comparing Figs.~\ref{fig:dlc} and
\ref{fig:dlc_wodnu} manifests that the neutrino perturbation does
contribute to the spherical collapse, which differs from the effect of
smooth dark energy model \cite{WangSteinhardt:98,Paceetal:10}.

\begin{figure}[t]
\begin{center}
\includegraphics[width=0.6\textwidth]{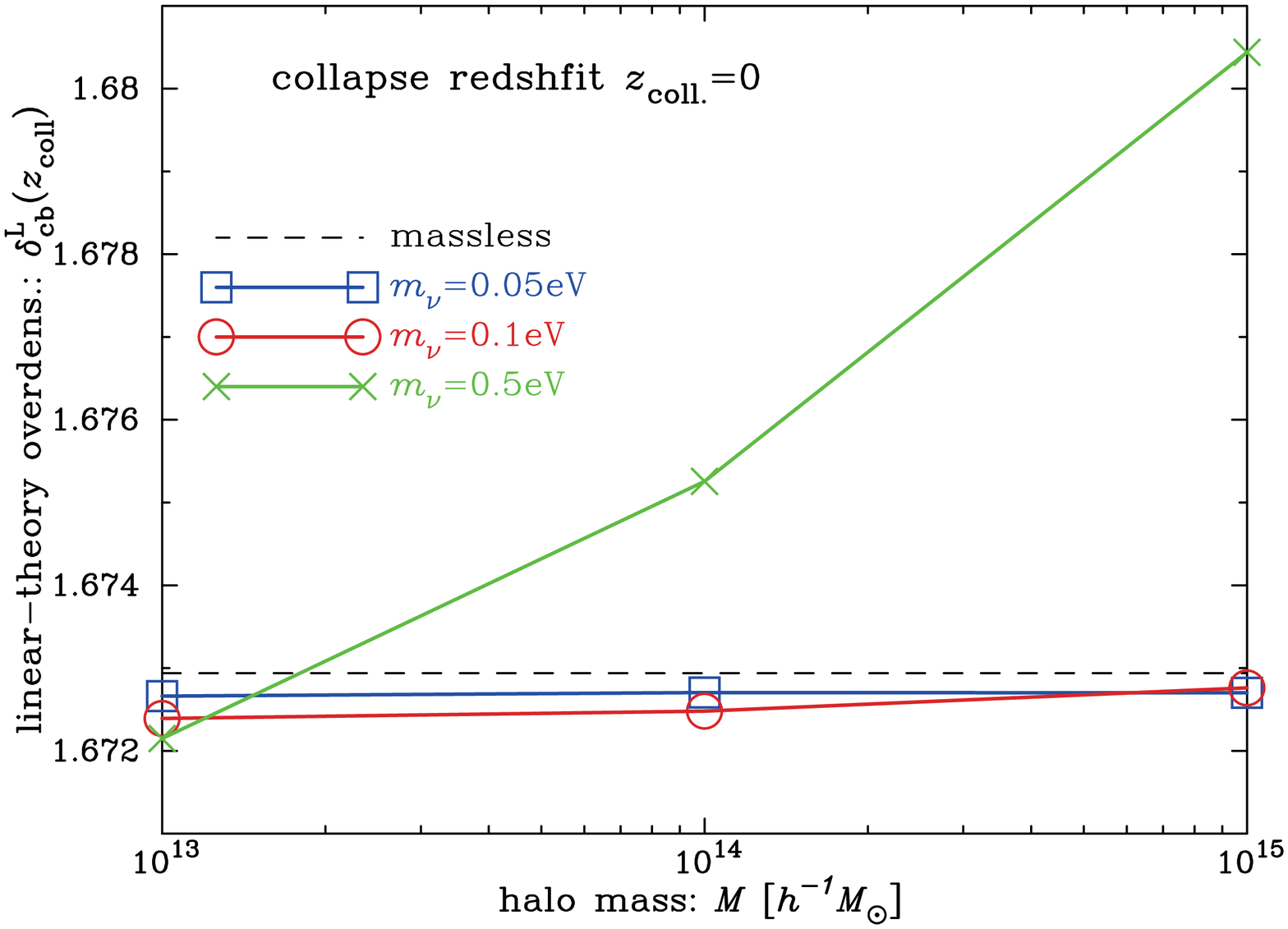}
\end{center}
\caption{The plot shows how the critical overdensity $\delta^{\rm
 L}_{cb}$ changes with different halo mass scales and neutrino mass
 scales. Here we consider $z_{\rm coll}=0$ for the collapse
 redshift. For comparison, the dashed curve shows the result without
 massive neutrino. More massive halos have a deeper gravitational
 potential well and therefore are more capable of capturing neutrinos
 around it. In addition, neutrinos of greater mass scales have a
 smaller free-streaming scale, and are more captured by the CDM top-hat
 region. These facts together with Figs.~\ref{fig:dlc} and \ref{fig:dlc_wodnu}
 explain the non-trivial dependence of $\delta_{cb}^{\rm L}$
 on halo mass scales as well as neutrino mass scales. 
}  \label{fig:massdependence}
\end{figure}

In Fig.~\ref{fig:massdependence}, we summarize dependences of the
critical overdensity on halo mass and neutrino mass, assuming the
collapse redshift $z_{\rm coll}=0$. It can be found that, for a fixed
halo mass scale, the critical overdensity first decreases with
increasing the neutrino mass from 0.05~eV, but then starts to increase
at greater neutrino mass scales from some mass scale. The turnover
neutrino mass scale slightly changes with halo mass scale. This
non-trivial dependence arises depending on which of the two competing
effects discussed above dominates.  If we ignore the neutrino
perturbation, the critical overdensity decreases with increasing
neutrino mass independently of halo mass.

Summarizing the results in Figs.~\ref{fig:dlc}, \ref{fig:dlc_wodnu} and
\ref{fig:massdependence}, we can conclude that the effect of massive
neutrino on the critical overdensity is very small, less than $\sim
0.5\%$, compared to the critical overdensity without massive neutrino,
for neutrino mass scales $m_\nu\simlt 0.5~{\rm eV}$ and halo mass scales
we are interested in.

\subsection{The impact of massive neutrinos on halo mass function}

In this subsection, we estimate the impact of massive neutrinos on the
halo mass function that is one of the most important observables for
cluster surveys. 

As we have shown, the effect of the massive neutrino on the nonlinear
gravitational collapse of CDM plus baryon perturbation is well captured by
the linear growth rate. In other words, the nonlinear neutrino
clustering around the CDM overdensity does not largely change the
nonlinear dynamics, and therefore is very unlikely to change structural
properties of mass distribution within a halo.  
We here assume that
the halo mass function for a MDM model can be obtained from a mapping of
the mass function in CDM models without massive neutrino. That is, we
assume that the mapping of halo mass function can be obtained by
assuming that (1) only the cold component (CDM plus baryon) can collapse
to form halos, and (2) the halo mass function for a MDM model can be
obtained just by replacing the linear-theory mass
fluctuations appearing in the mass function for a CDM model with the
corresponding mass fluctuation of CDM plus baryon perturbation for a MDM
model:
\begin{equation}
\frac{dn}{d\ln M}(z)=\frac{\bar{\rho}_{cb}}{M} \times f\!\!\left( \nu=
\frac{\delta_{cb,{\rm crit}}^L(M; z)}{\sigma_{bc}(M; z)} \right) \frac{d\nu}{d\ln
M}, \label{eq:dndlnM}
\end{equation}
where the function $f(\nu)$ is the fitting formula that is obtained
based on a suit of N-body simulations for CDM models. The previous works
have shown that the fitting formula is well characterized in terms of
the peak height, $\nu\equiv \delta^{\rm L}_{\rm crit}/\sigma(M;z)$
\cite{ShethTormen:99,Tinkeretal:08}, where $\delta_{\rm crit}^{\rm L}$
is the linear-theory extrapolated critical overdensity for halo
formation at a given redshift and $\sigma(M; z)$ is the linear rms mass
fluctuation smoothed with the halo mass scale $M$ and at redshift
$z$. In Eq.~(\ref{eq:dndlnM}), we assumed that we can obtain the halo
mass function for a MDM model simply by using the peak height for CDM
plus baryon perturbation as well as by using the prefactor
$\bar{\rho}_{cb}/M$, the mean mass density of CDM plus baryon, because
the cold component is the collapsing component to form halos. To be more
precise, the rms mass fluctuation of halo mass $M$ for CDM plus baryon
perturbation is defined as
\begin{equation}
\sigma^2_{cb}(M;z)\equiv
 \int_0^{\infty}\!\frac{dk}{k}\frac{k^3}{2\pi^2}P_{cb}^{\rm
 L}(k;z)\tilde{W}^2(kR_M), 
\label{eq:sigmam_cb}
\end{equation}
where $P^{\rm L}_{cb}(k;z)$ is the linear power spectrum of CDM plus
baryon perturbation at target redshift $z$, and $\tilde{W}(kR_M)$ the
Fourier-transformed top-hat filter: $\tilde{W}(x)\equiv 3(\sin x-x\cos
x)/x^3$. The filtering scale and  halo mass are related via 
$M=(4\pi/3)\bar{\rho}_{cb,0} R_M^3 $ ($\bar{\rho}_{cb, 0}$ is the
present-day mean mass density of CDM plus baryon). 

As for the fitting formula, we use the formula in
\cite{Bhattacharyaetal:11} that is obtained from N-body simulations for
a range of CDM models varying around the fiducial cosmological model
consistent with the WMAP data:
\begin{equation}
f(\nu,z)=A\sqrt{\frac{2}{\pi}}\exp\left[
-\frac{a\nu^2}{2}
\right]\left[1+\left(a\nu^2\right)^p\right]
\left(\sqrt{a}\nu\right)^q \frac{1}{\nu},
\end{equation}
where $A=0.333(1+z)^{-0.11}$, $a=0.788(1+z)^{-0.01}$, $p=0.807$ and
$q=1.795$. For the peak height $\nu=\delta_{\rm crit}/\sigma(M;z)$,
\cite{Bhattacharyaetal:11} simply used the fixed critical overdensity
$\delta_{\rm crit}=1.686$, the value of Einstein de-Sitter model, 
 and then found the best-fit parameters $A, a$
and so on by fitting the functional form above with the mass function
measured from simulations for variant CDM models. 
More exactly speaking, when we have the effects of baryon perturbation
and cosmic acceleration, the critical overdensity $\delta^{\rm L}$ is
changed from the Einstein de-Sitter value $\delta^{\rm
L}=1.686$. However, the change is very small, less than a percent level
(see Fig.~\ref{fig:dlc}), and therefore it was assumed that the change of
$\delta^{\rm L}$ is absorbed by tuning the fitting model parameters.
If the change of $\delta^{\rm
L}$ is properly taken into account, the fitting will yield slightly different
best-fit model parameters  of
$A, a$ and so on. Furthermore, although the presence of baryon
perturbation changes the collapse of CDM perturbation (see
Fig.~\ref{fig:compare_womnu}), 
we here assume that the
simulations in \cite{Bhattacharyaetal:11} properly take into account
the effect of baryon when setting up the initial conditions of N-body
simulations (see below for a further discussion).
To compute $\sigma_{cb}(M;z)$ in Eq.~(\ref{eq:dndlnM}),
we use the CAMB code \cite{CAMB} to
compute the transfer functions. The CAMB outputs include
the effect of massive neutrinos or baryon perturbations on the
growth of CDM perturbation.

\begin{figure}[t]
\begin{center}
\includegraphics[width=0.6\textwidth]{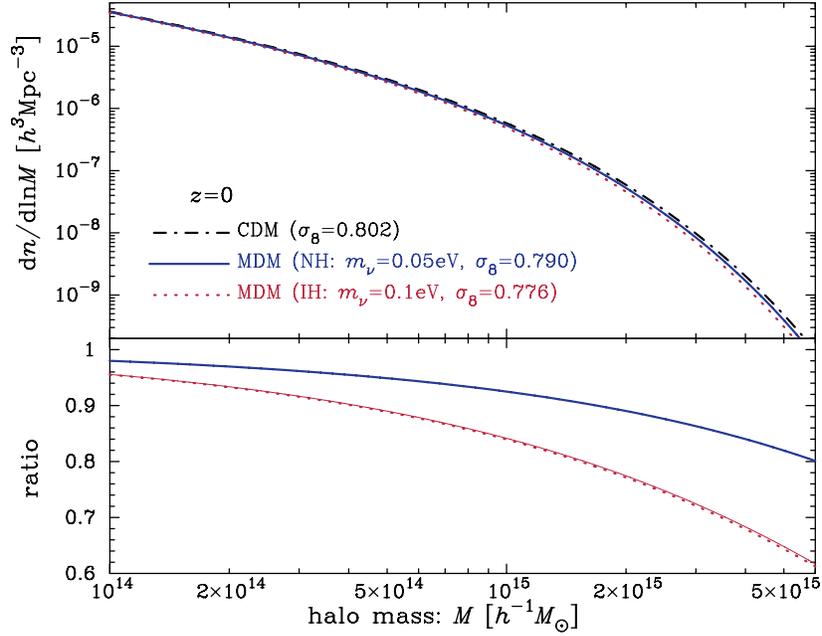}
\end{center}
\vspace*{-2em}
\caption{The upper panel shows the halo mass function at $z=0$ for
 CDM-dominated models with and without massive neutrinos. The halo mass
 function for a mixed dark matter model (CDM plus massive neutrino) is
 computed by mapping the fitting formula for CDM model based simulations
 using  Eq.~(\ref{eq:dndlnM}). For the solid and dotted curves, we
 assume the neutrino mass scales $m_\nu=0.05$ and 0.1~eV, which are
 close to the lower mass bounds for the normal and inverted mass
 hierarchies (NH and IH), respectively. The presence of massive
 neutrino, for a fixed $\Omega_{\rm c0}+\Omega_{\nu0}$, decreases the abundance of
 massive halos. The lower panel explicitly shows the ratio of the mass
 functions for models with and without massive neutrino
 contribution. The linear mass fluctuation such as $\sigma_8$ changes
 only by a few percent at most for these neutrino masses, however, the
 abundance of massive halos may decrease by up to a 
factor 2
at a few
 $10^{15}h^{-1}M_\odot $.}
 \label{fig:dndm}
\end{figure}

\begin{figure}
\begin{center}
\includegraphics[width=0.95\textwidth]{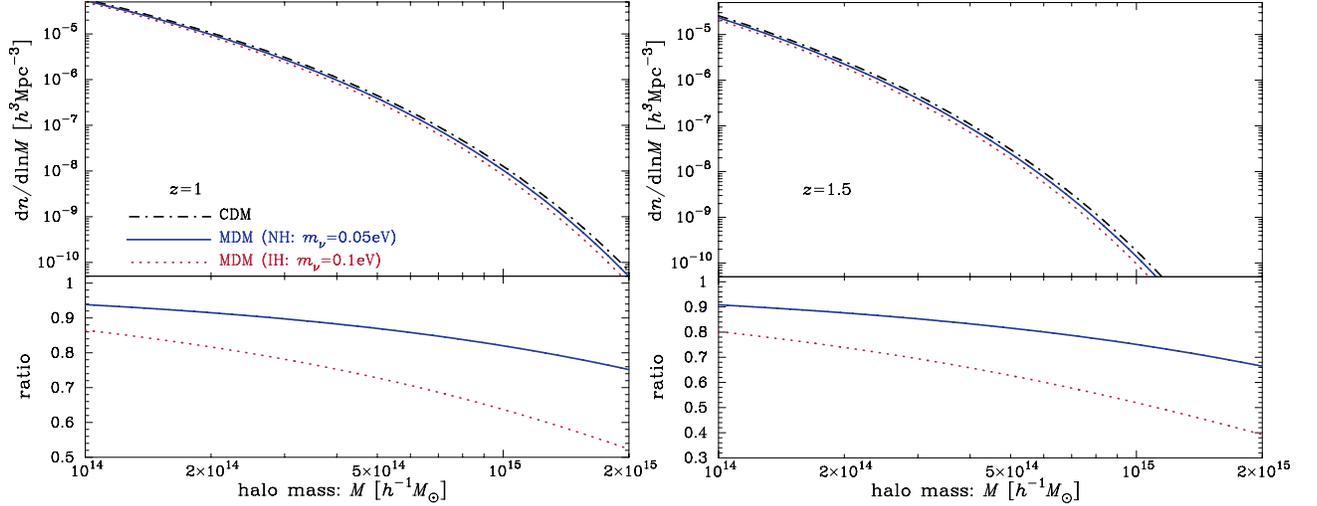}
\end{center}
\vspace*{-2em}
\caption{Similarly to the previous plot, but for $z=1$ (left) and 1.5
 (right panel), respectively. The effect of massive neutrino on the
 abundance of halos for a fixed halo mass scale is more significant at
 higher redshift.}
\label{fig:dndm_z1}
\end{figure}

The upper panel of Fig.~\ref{fig:dndm} compares the halo mass functions
at $z=0$ for different models with and without massive neutrino
contribution. 
The lower panel explicitly shows the ratio of the mass functions with
and without massive neutrino. Here we assumed $m_\nu=0.05$ and $0.1$~eV
for the neutrino mass scale, which are close to the lower bounds of the
normal mass hierarchy (NH) and the inverted hierarchy (IH) that are
implied from the terrestrial experiments. Hence either of these results
would inevitably exist in our universe. The presence of massive neutrino
decreases the abundance of halos, more significantly for more massive
halos that reside in the exponential tail of mass function. The decrease
in the halo abundance is up to a factor 2 around $\sim 5\times
10^{15}~h^{-1}M_\odot$. This change can be compared to the effect on the
linear mass fluctuation such as $\sigma_8$; the neutrino of these mass
scales decreases $\sigma_8$ only by a few percent for neutrinos of these
mass scales.  Again the higher sensitivity of halo mass function to
neutrino mass is through the exponential tail of mass function at
massive halo ends. The thin solid curve in the lower panel (although
almost overlapped with the dotted curve) shows the ratio when further
taking into account the change in the critical density $\delta^{\rm
L}_{\rm cb}$ in the mass function (Eq.~\ref{eq:dndlnM}); more explicitly
we decreases the critical density by 0.03\%, a maximum change implied
from Fig.~\ref{fig:massdependence} for the case of $m_{\nu,{\rm
tot}}=0.1~$eV. It is clear that the change in the critical density due
to the massive neutrinos causes a negligible effect on the halo mass
function.

Fig.~\ref{fig:dndm_z1} shows the similar results, but for higher
redshifts $z=1$ and $1.5$, respectively. The decrease in the 
abundance of cluster scale-halos is more significant at higher
redshifts. 

\begin{figure}
\begin{center}
\includegraphics[width=0.6\textwidth]{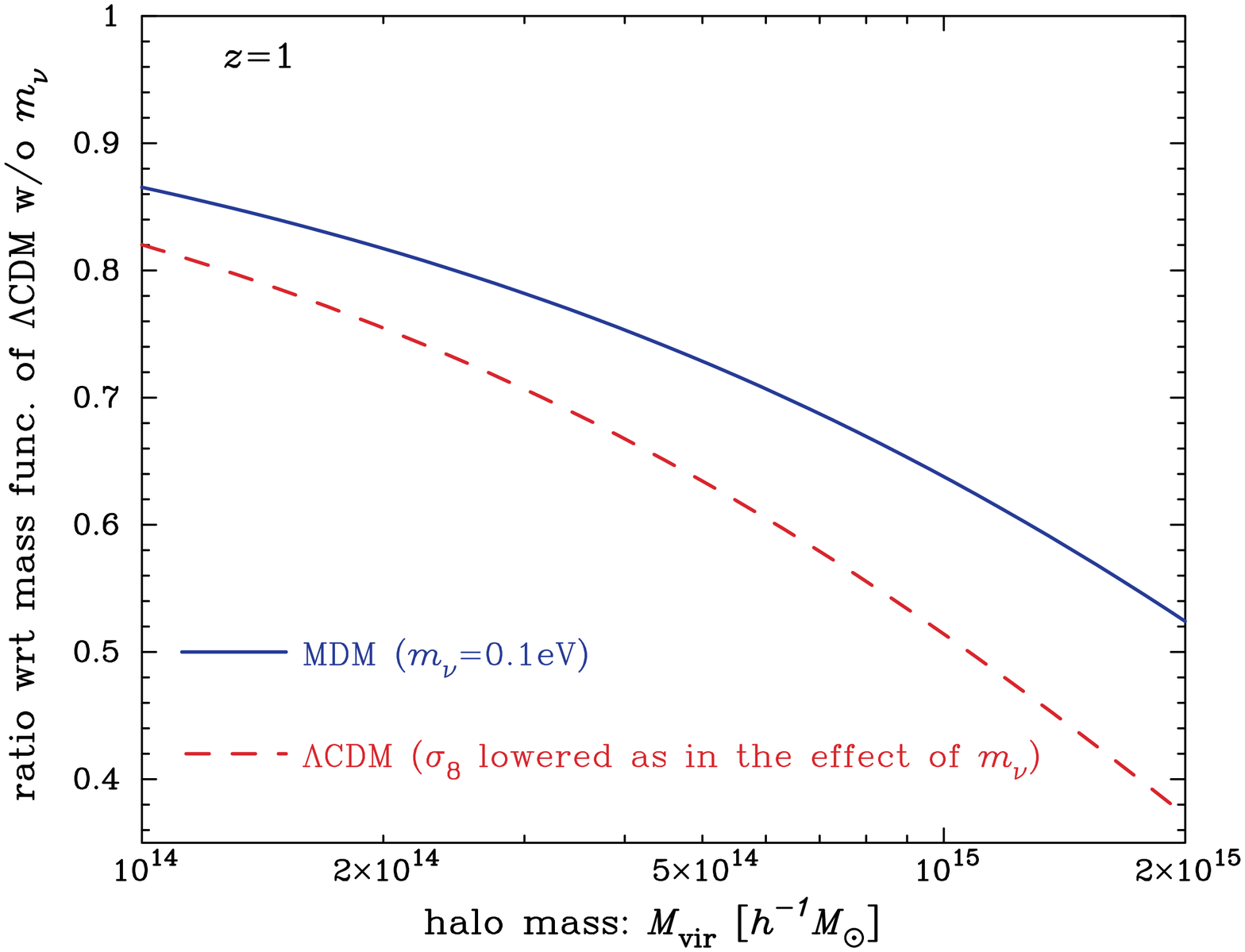}
\end{center}
\vspace*{-2em}
\caption{Similar plot to the lower panel in the previous figure, but we
 here compare the halo mass function for a MDM model with $m_\nu=0.1~$eV
 to the mass function for a CDM model (without massive neutrino), where
 $\sigma_8$ value is lowered by the same amount as in the neutrino
 suppression effect on the linear mass fluctuation at $8~h^{-1}$Mpc for
 the MDM model. More exactly, the $\sigma_8 $ value at $z=1$ is changed
 to 0.486 from 0.500 for the dashed curve. The CDM model with
 normalization of the lowered $\sigma_8$ value reproduces
 the mass function for the MDM model, for the same $\Omega_{\rm
 c0}+\Omega_{\nu 0}$, within 30\% level accuracy over the range of halo
 masses we consider.  } \label{fig:dndm_sig8}
\end{figure}

One may think whether or not the effect of massive neutrino on the halo
mass function is mostly described by the change of $\sigma_8$, the
normalization parameter of power spectrum amplitudes often used in the
literature. Fig.~\ref{fig:dndm_sig8} compares the halo mass functions
for a MDM model with $m_\nu=0.1~$eV and for a $\Lambda$CDM model where
$\sigma_8$ at $z=1$ is lowered so as to match the $\sigma_8$ value
for the MDM model; more precisely, $\sigma_8(z=1)$ is changed to 0.486 
from 0.500. Note that both the models have the same $\Omega_{\rm c0}+\Omega_{\nu0}$. 
The $\Lambda$CDM model with the lowered $\sigma_8$ 
roughly reproduces the decrease in the halo abundance. 
However, the two curves do not exactly agree because of the difference
in the linear power spectra of CDM and baryon perturbations.
Nevertheless, 
this gives a justification of the neutrino mass constraint derived in
\cite{Vikhlininetal:09}, where the neutrino mass constraint is obtained
from the allowed range of $\sigma_8$ values that are derived by
comparing the observed abundance of X-ray luminous clusters with the
model halo mass functions varying within CDM models without massive
neutrino contribution.

\subsection{Discussion: Cosmological parameter degeneracies}

\begin{figure}
\begin{center}
\includegraphics[width=17cm,height=7cm]{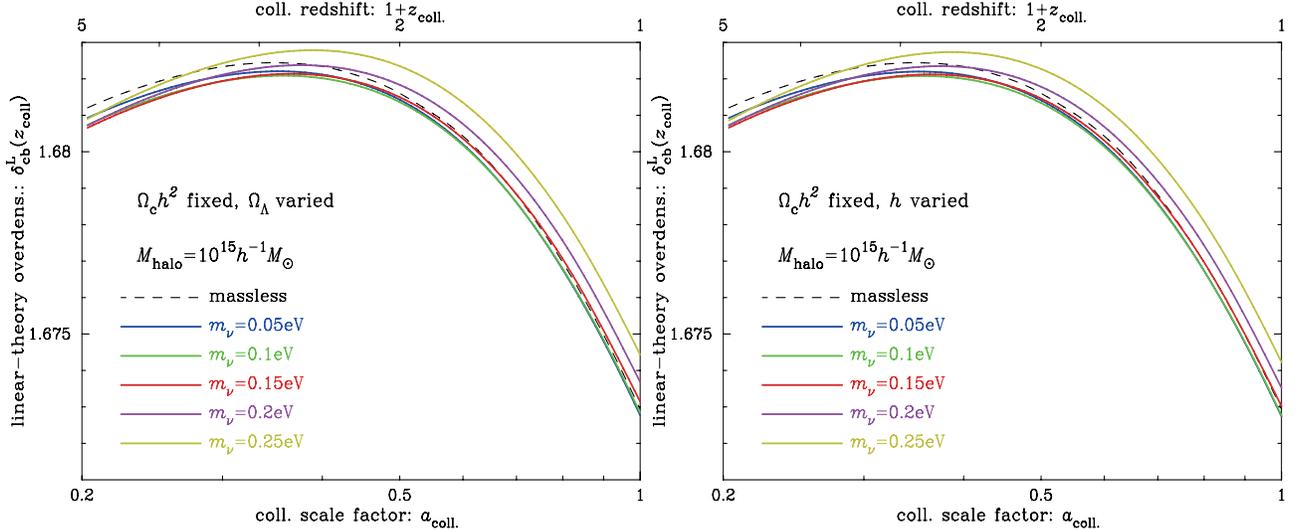}
\end{center}
\vspace*{-2em}
\caption{ As in Fig.~\ref{fig:dlc}, the linear-theory extrapolated
critical density for halos of $M=10^{15}h^{-1}M_\odot $ as a function of
collapse redshift for MDM models of different neutrino mass scales,
where we added massive neutrinos around the fiducial $\Lambda$CDM model
by changing either $h$ or $\Omega_\Lambda$ parameter with fixing the CDM
density parameter $\Omega_{\rm c0}h^2$ and keeping the flat geometry
$\Omega_{\rm c0}+\Omega_{\rm b0}+\Omega_{\nu0}+\Omega_\Lambda=1$ (see
text for details).  Note that, for the previous plots, we varied the CDM
dark matter density $\Omega_{\rm c0}$ with fixing the total dark matter
density $\Omega_{\rm c0}+\Omega_{\nu0}$ to the fiducial value when
adding massive neutrinos.  } \label{fig:delta_c_och2fixed}
\end{figure}

\begin{figure}
\begin{center}
\includegraphics[width=0.6\textwidth]{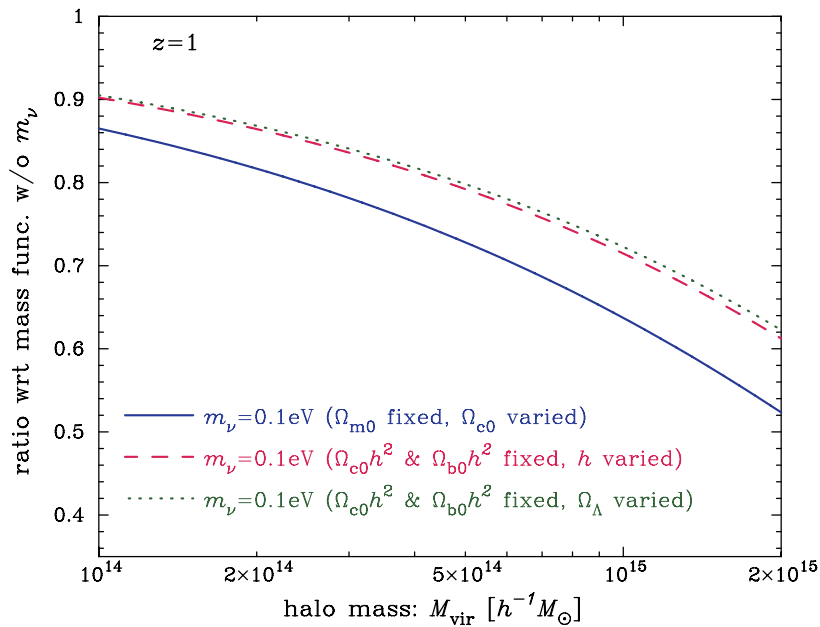}
\end{center}
\vspace*{-2em}
\caption{Shown is the ratio of the halo mass functions with and without
 massive neutrinos of $m_{\nu,{\rm tot}}=0.1~$eV. The different curves
 show the results when changing either $h$, $\Omega_\Lambda$ or
 $\Omega_{\rm c0}$ alone around our fiducial cosmological model (see the
 end of Sec.~\ref{sec:intro}) with fixing the parameters $\Omega_{\rm
 b0}h^2$ and assuming a flat geometry of $\Omega_{\rm c0}+\Omega_{\rm
 b0}+\Omega_{\nu0}+\Omega_\Lambda=1$.  We have so far considered the
 case varying $\Omega_{\rm c0}$, and the cases varying $h$ or
 $\Omega_\Lambda$ are motivated by the fact that the CMB observables
 well constrain the curvature parameter, $\Omega_{\rm c}h^2$ and
 $\Omega_{\rm b0}h^2$.  Around our fiducial cosmological model, these
 are equivalent to the parameter changes, $h=0.71\rightarrow h=0.7128$,
 $\Omega_{\Lambda}=0.7354 \rightarrow 0.7333$ or $\Omega_{\rm c0}=0.2198
 \rightarrow 0.2177$, respectively.  } \label{fig:fm_h+Ol}
\end{figure}

When adding the massive neutrinos for
different masses, we have so far kept the total dark matter density,
$\Omega_{\rm c0}+\Omega_{\nu0}$, fixed. For a more practical
perspective, the CMB information give precise constraints on the CDM and
baryon densities, $\Omega_{\rm c0}h^2$ and $\Omega_{\rm b0}h^2$, as well
as the curvature parameter or equivalently the total energy density,
$\Omega_{\rm c0}+\Omega_{\rm b0}+\Omega_{\nu0}+\Omega_{\Lambda}\simeq
1$. Massive neutrinos with small mass scales of a few $\simlt 0.1~$eV
were relativistic before the decoupling epoch, and do not affect the CMB
observables. Therefore the CMB observables cannot well constrain the
neutrino mass of the small mass scales, leaving degeneracies in
cosmological parameters. Given these facts one might think that, when
adding the massive neutrinos, we should keep these CMB-constrained
parameters fixed. If we assume a flat geometry, this is equivalent to
varying either the Hubble parameter $h$ or the energy density of the
cosmological constant $\Omega_{\Lambda}$ with fixing the CMB parameters
above. For example, when the neutrino mass $m_{\nu, {\rm tot}}=0.1~$eV
is added, this leads to $h=0.7128$ or
$\Omega_\Lambda=0.7333$ from the fiducial values $h=0.71$ or
$\Omega_\Lambda=0.7354$, respectively.

Fig.~\ref{fig:delta_c_och2fixed} shows the critical density for
halos of $M=15^{15}h^{-1}M_\odot$ for a MDM model with various neutrino
mass scales, where we varied either $\Omega_{\Lambda}$ or $h$ by the
amount determined by the neutrino mass scale, but fixing $\Omega_{\rm
c0}h^2$. The results are similar to Fig.~\ref{fig:dlc}; the effect of
massive neutrinos on the critical density is very small for the range of
cosmological models.  Fig.~\ref{fig:fm_h+Ol} shows how the MDM models
alter the halo mass function compared to the case without
massive neutrinos, which can be compared with our fiducial case where
the massive neutrinos of $m_{\nu,{\rm tot}}=0.1~$eV are added by varying
the CDM density parameter $\Omega_{\rm c0}$. The parameter change of $h$
or $\Omega_\Lambda$ also leads to the smaller abundance of massive halos
as in the case changing $\Omega_{\rm c0}$, but the decrease is slightly
smaller than the case when changing $\Omega_{\rm c0}$.

\section{Summary}

In this paper, we have developed a method to solve the nonlinear
dynamics of top-hat CDM overdensity region including the effects of
baryon perturbation and massive neutrinos.  In developing the spherical
collapse model, we properly set up the initial conditions of each
components (baryon, CDM and neutrinos), which have different amplitudes
and profiles, assuming the adiabatic initial conditions (see
Fig.~\ref{fig:sp_collapse}).  In fact we found that the nonlinear
dynamics is very sensitive to detailed setup of the initial conditions
of top-hat CDM perturbation, more precisely $\delta_c(z_i)$ and the
velocity of top-hat radius $\dot{R}(z_i)$. 
For example, we
cannot employ the linear-theory prediction for an Einstein de-Sitter
model, $\delta\propto a$, to set up the initial conditions, e.g., even
at an epoch in the sufficiently linear regime such as the decoupling
epoch $z_{\rm ini}\simeq 1100$, because this solution ignores that the
CDM perturbation is affected by the presence of baryon and
massive neutrino.

Since we cannot treat the neutrinos as a perfect fluid, we properly
solved the linearized Boltzmann hierarchy equations to compute time
evolution of linearized neutrino perturbations, where we include the effect of
nonlinear gravitational potential due to the nonlinear CDM and baryon
perturbation in the late stage. For neutrino mass scales lighter than a
few $0.1$~eV, the range inferred from the neutrino oscillation
experiments and the cosmological constraints, the neutrino perturbation
stays in the quasi nonlinear regime, $\delta_\nu\simlt 1$ (see
Fig.~\ref{fig:density}). This gives a justification of our treatment
where we used the {\em linearized } Boltzmann equations. As for an
improved modeling, one can further include the nonlinear terms such as
the coupling term between the nonlinear gravitational potential and the
perturbed phase-space density of neutrinos in order to solve the time
evolution of neutrino perturbations in a perturbation theory manner.

By solving the spherical collapse model for cosmological models around a
$\Lambda$CDM model that is consistent with the WMAP data, we found that
both the neutrino and baryon perturbations delay the collapse of CDM
overdensity compared to a model with CDM alone
(Fig.~\ref{fig:compare_womnu}). However, interestingly we found that the
collapse redshift can be well monitored by the linear-theory
extrapolated overdensity of CDM (plus baryon) perturbation(s) for the
ranges of neutrino masses ($\simlt $ a few 0.1~eV) and halo mass scales
we have considered. This result is promising because the linear-theory
extrapolated overdensity (the critical density) can be accurately
computed using the linear perturbation theory, once cosmological model
and neutrino mass are specified.  In other words, we found that the
massive neutrinos with the rang of mass scales lead to only a small
change in the critical density
by $\lesssim 0.1$\% compared to the model without massive neutrino, but
with the same $\Omega_{\rm m0}$ (Figs.~\ref{fig:dlc}, \ref{fig:dlc_wodnu}, and
\ref{fig:massdependence}).

Given the results of the spherical collapse model, we gave
Eq.~(\ref{eq:dndlnM}) to estimate the halo mass function including the
effect of massive neutrinos, where the effect of massive neutrinos are
properly taken into account in the linear mass fluctuations of CDM and
baryon perturbations at a given redshift, smoothed with a given halo
mass scale; $\sigma_{cb}(M,z)$ \citep[also see][for the similar
discussion]{Brandbygeetal:10}. Using the equation, we found that the
presence of massive neutrinos with $0.05$ and 0.1~eV, the lower-bound
mass scales of normal and inverted mass hierarchies, respectively, may
cause a significant decrease in the abundance of massive halos; more
specifically, up to a factor of 2 for halos with $10^{15}M_\odot$ and at
$z\sim 1$ (see Figs.~\ref{fig:dndm} and \ref{fig:dndm_z1}). Thus our
results imply that massive neutrinos, which {\em should} exist in our
universe, relax to some extent a possible tension that the cutting-edge
SZ experiments could not find as many massive clusters as what was
originally expected \cite{ACT:11,SPT:11}. This needs to be further
studied more carefully. Since it is still
challenging to accurately simulate nonlinear structure formation in a
MDM model, especially for such light neutrino mass scales of $\lesssim$
a few $0.1~$eV \citep[see][for the
attempts]{BrandbygeHannestad:09,Brandbygeetal:10,Vieletal:10}, the
analytical model developed in this paper will give a useful tool or at
least useful guidance for interpreting ongoing and upcoming wide-area
surveys of massive clusters.

Our findings also propose several applications. First, as we stressed
above, a careful setup of the initial conditions is very important in
order to have an accurate nonlinear dynamics, for  a
multi-component system with CDM, baryon and neutrinos. This implies that
it is very important to set up the accurate initial conditions for
cosmological simulations including the effect of baryon such as 
smoothed particle hydrodynamical (SPH) simulations \citep[see][for the
similar
discussion]{Yoshidaetal:03,TseliakhovichHirata:10,Naozetal:11}. Since
the spherical collapse model gives an exact solution of the nonlinear
dynamics, albeit an unrealistic symmetry assumed, we can explore how to
set up the initial conditions by combining the spherical collapse model
with the linear and/or perturbation theory predictions. For example, it
was shown that using the second-order Lagrangian perturbation theory
allows one to set up more accurate initial conditions of N-body
simulations that are simulations for a model with CDM-alone or single
cold component \cite{ScoccimarroSheth:02,Jenkins:10}. We can extend this
analysis to a multi-component system; we can apply the second-order
Lagrangian perturbation theory to CDM and baryon perturbations
separately by taking into account the different growth rates, and then
can study how the improved initial conditions can reproduce the exact
solution of spherical collapse for CDM and baryon perturbations starting
from a given initial redshift. Such a study will give a useful guidance
for exploring how to set up the initial conditions for CDM and baryon
particles in a SPH-type simulation. 
This can be further extended to a case
further including the neutrino particles. These are our future
study, and will be presented elsewhere.

Secondly, several studies recently claimed that detected massive
clusters at high redshifts beyond $z\sim 1$ may give a tension of
$\Lambda$CDM structure formation model \citep[][also see references
therein]{Jeeetal:09,HolzPerlmutter:10}. It is indeed interesting to
explore whether or not these particular catalogs of clusters, which are
found by different observations/surveys under different selection
functions, can falsify the $\Lambda$CDM predictions as explored in
\cite{Mortonsonetal:11,SPT:11}. However, the effect of massive neutrinos
has been ignored in the previous studies. Again, the method developed in
this paper can be used to address how the presence of the high-$z$
massive clusters may falsify a more realistic cosmological model that includes
massive neutrino contribution. This study will be presented elsewhere.

\bigskip
\noindent{\bf Acknowledgement}: We thank Scott Dodelson, 
Salman Habib, 
Wayne Hu, 
Chung-Pei Ma, Daisuke Nagai, Smadar Naoz, Shun Saito, 
Roman Scoccimarro, Tristan Smith,
and David Spergel
for useful
discussion.  This work is supported in part by the Grant-in-Aid for the
Scientific Research Fund (Nos. 21740177 and 23340061), 
by JSPS
Core-to-Core Program ``International Research Network for Dark
Energy'',
by World Premier
International Research Center Initiative (WPI Initiative), MEXT, Japan, 
and by the FIRST program
``Subaru Measurements of Images and Redshifts (SuMIRe)'', CSTP, Japan.

\bibliographystyle{apsrev}
\bibliography{nu}

\end{document}